\documentclass[preprint,journal]{vgtc}       % preprint (journal style)

\ifpdf%                                % if we use pdflatex
  \pdfoutput=1\relax                   % create PDFs from pdfLaTeX
  \usepackage{graphicx}                % allow us to embed graphics files
  \DeclareGraphicsExtensions{.pdf,.jpg,.jpg,.jpeg} % for pdflatex we expect .pdf, .jpg, or .jpg files
\else%                                 % else we use pure latex
  \usepackage{graphicx}                % allow us to embed graphics files
  \DeclareGraphicsExtensions{.eps}     % for pure latex we expect eps files
\fi%

\graphicspath{{figures/}{pictures/}{images/}{./}} % where to search for the images

\usepackage{microtype}                 % use micro-typography (slightly more compact, better to read)
\PassOptionsToPackage{warn}{textcomp}  % to address font issues with \textrightarrow
\usepackage{textcomp}                  % use better special symbols
\usepackage{mathptmx}                  % use matching math font
\usepackage{times}                     % we use Times as the main font
\usepackage{cite}                      % needed to automatically sort the references
\usepackage{tabu}                      % only used for the table example
\usepackage{booktabs}                  % only used for the table example
\usepackage{amsmath}
\usepackage{wrapfig}
\usepackage[bookmarks=false]{hyperref}

\usepackage{color}
\newcommand{\revise}[2]{\textcolor{black}{#2}}

\onlineid{1159}

\vgtccategory{Research}
\vgtcpapertype{algorithm/technique}

\title{IlluminatedFocus: Vision Augmentation using Spatial Defocusing via Focal Sweep Eyeglasses and High-Speed Projector}
\author{Tatsuyuki Ueda, Daisuke Iwai, \textit{Member, IEEE}, Takefumi Hiraki, \textit{Member, IEEE}, and Kosuke Sato, \textit{Member, IEEE}}
\authorfooter{
\item
 Daisuke Iwai is with Osaka University and JST, PRESTO. E-mail: daisuke.iwai@sys.es.osaka-u.ac.jp.
\item
 Tatsuyuki Ueda, Takefumi Hiraki, and Kosuke Sato are with Osaka University.
}

\shortauthortitle{Ueda \MakeLowercase{\textit{et al.}}: IlluminatedFocus: Vision Augmentation using Spatial Defocusing via Focal Sweep Eyeglasses and High-Speed Projection}
\abstract{
Aiming at realizing novel vision augmentation experiences, this paper proposes the IlluminatedFocus technique, which spatially defocuses real-world appearances regardless of the distance from the user's eyes to observed real objects.
With the proposed technique, a part of a real object in an image appears blurred, while the fine details of the other part at the same distance remain visible.
We apply \revise{electrically focus tunable lenses}{Electrically Focus-Tunable Lenses} (ETL) as eyeglasses and a synchronized high-speed projector as illumination for a real scene.
We periodically modulate the focal lengths of the glasses (focal sweep) at more than 60 Hz so that a wearer cannot perceive the modulation.
A part of the scene to appear focused is illuminated by the projector when it is in focus of the user's eyes, while another part to appear blurred is illuminated when it is out of the focus.
As the basis of our spatial focus control, we build mathematical models to predict the range of distance from the ETL within which real objects become blurred on the retina of a user.
Based on the blur range, we discuss a design guideline for effective illumination timing and focal sweep range.
We also model the apparent size of a real scene altered by the focal length modulation.
This leads to an undesirable visible seam between focused and blurred areas.
We solve this unique problem by gradually blending the two areas.
Finally, we demonstrate the feasibility of our proposal by implementing various vision augmentation applications.
} % end of abstract


\CCScatlist{ % not used in journal version
 \CCScat{K.6.1}{Management of Computing and Information Systems}%
{Project and People Management}{Life Cycle};
 \CCScat{K.7.m}{The Computing Profession}{Miscellaneous}{Ethics}
}

\teaser{
  \centering
  \includegraphics[width=\linewidth]{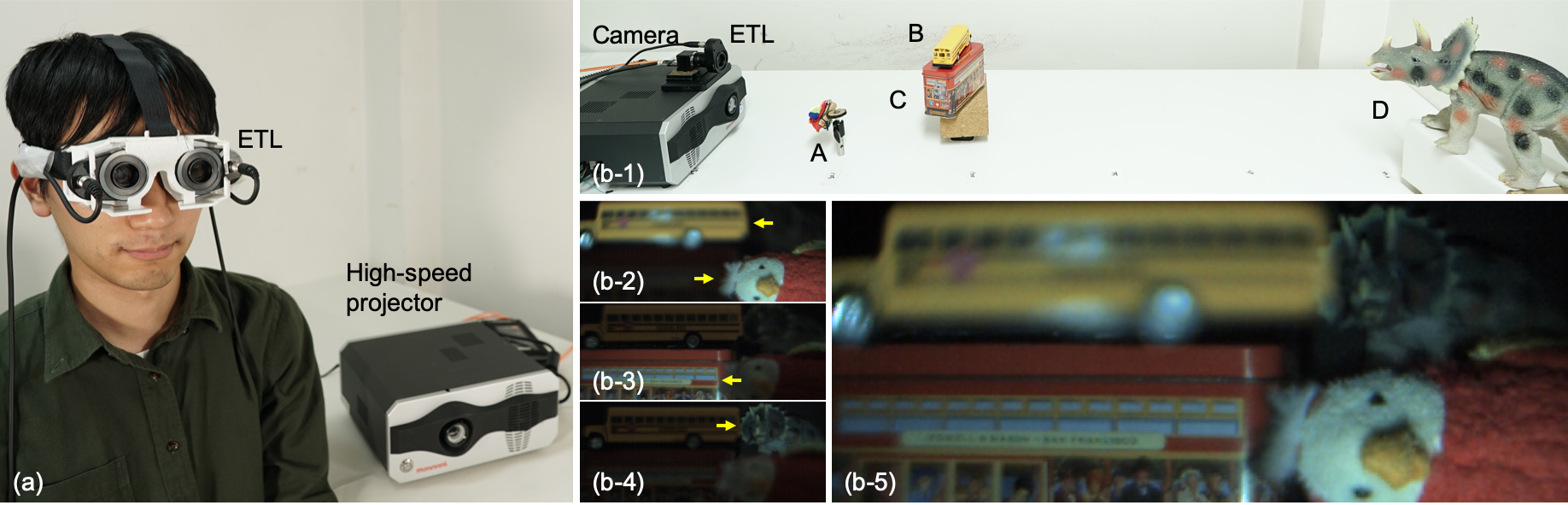}
  \caption{The IlluminatedFocus technique optically defocuses real-world appearances in a spatially varying manner regardless of the distance from the user's eyes to observed real objects. The proposed technique enables various vision augmentation applications. \revise{}{(a) The proposed system consists of focal sweep eyeglasses (two Electrically Focus-Tunable Lenses (ETL)) and a high-speed projector. (b) Experimental proof of the proposed technique. (b-1) Experimental setup. Four objects (A, B, C, and D) are placed in front of the projector and ETL. A camera is regarded as a user's eye. (b-2, b-3, b-4) The objects are illuminated by the projector at different timings and the camera's focal length is periodically modulated by the ETL. As indicated by the yellow arrows, objects to appear focused (A, C, and D) are illuminated when they are in focus and the other object to appear blurred (B) is illuminated when it is out of focus. (b-5) When the frequency of the focal sweep is higher than the critical fusion frequency (CFF), these appearances are perceived to be integrated. The appearance of this image (only B is blurred) cannot be achieved by normal lens systems.} Note that the brightness of (b-2) to (b-5) has been adjusted for better understanding.}
	\label{fig:teaser}
}

\vgtcinsertpkg

\begin{document}


\firstsection{Introduction}

\maketitle

% 1
% ボケはいろいろ使える
Focusing and defocusing are important optical effects for human vision to understand the three-dimensional (3D) structure of a real scene.
People with normal vision can perceive the fine details of an object at which they gaze, while the details of other objects beyond the depth-of-field (DOF) are lost due to defocus blur.
%Leveraging this characteristics, t
Various graphical user interfaces (GUI) (e.g., Launchpad in MacOS and background blur in Skype) have applied a blur effect because it naturally decreases visual clutter of a background and consequently makes the foreground information more comprehensible. %and consequently draws a user's attention to the other areas.
In addition to these examples, the blur effect realizes a wide range of fundamental human-computer interaction (HCI) techniques, such as visual guidance \cite{doi:10.1080/17470218.2012.722659,Hata:2016:VGU:2909132.2909254,mti3010019},
focus and context (F+C) visualization \cite{963286,974515,doi:10.1111/cgf.12877},
concealing undesired visual information (e.g., privacy protection) \cite{Yao:2013:FMA:2491367.2491377,10.1007/978-3-319-25554-5_25,8014910},
and enhancing the depth perception of a displayed image \cite{4480749,okatani,10.1007/978-3-642-23834-5_1,Held:2010:UBA:1731047.1731057}.
These interaction techniques are essential in augmented reality (AR) and vision augmentation applications, too.
Vision augmentation is a concept that enhances our vision using dynamic optics.
To deploy them in these systems, spatial focus control of a real-world scene is required.

% 2
% ビデオシースルーだと簡単に実装できるが、視野が狭いとかのdisadvantageもある。
% 光学的に実装できれば自然であるが、defocus blurは物理的にレンズからの距離にのみ依存するので、同じdepthにある対象の一部だけをぼかすとかは、一般的なレンズシステムでは物理的に不可能 (pSLM使った方法とかで可能だが空間解像度が低すぎる)
% プロジェクションマッピングで実世界を描きかえることも可能だが、解像度が足りない [mihara, anselm tip]
% fine details, edges of the scene cannot be visually canceled.
Video see-through (VST) AR systems can relatively easily blur a real scene.
The scene is captured as a digital image and displayed with superimposed virtual objects on a VST display.
Blurred real scenes can be synthesized simply by computing the convolution of a point spread function (PSF) and the captured image.
We can control the blur intensity over the image by preparing a spatially-varying PSF map.
\revise{On the other hand, spatial blur manipulation is not trivial in both optical see-through (OST) and spatial AR (SAR) systems, which are the platforms of vision augmentation.}{Although VST-AR can realize blur-based interfaces, it has a drawback that the real-world appearance is displayed with a perceivable latency and a limited field-of-view (FOV). On the other hand, both optical see-through (OST) and spatial AR (SAR) systems do not suffer from the latency and FOV problems and have been used as the platforms of vision augmentation, while spatial blur manipulation is not trivial in these systems.}
In OST-AR, typical displays do not have any focus control mechanisms for real-world viewing.
%-------英文校正ここから
Recently, an OST display that supports a mechanism to correct the focus of real-world objects has been developed \cite{8458263}.
However, defocus blur is depth-dependent; thus, it is physically impossible to spatially control defocus blurs regardless of the distance between the display and observed real objects.
Recent advances in the optics of virtual reality (VR) head-mounted displays (HMD) have achieved spatial focus control for virtual imagery using a free-form lens or a phase-only spatial light modulator \cite{8642529,Matsuda:2017:FSD:3072959.3073590}.
Although these techniques have a potential to achieve the spatial defocusing of a real-world scene, the spatial resolution of the focal length modulation is theoretically too low to support various interaction techniques.
\revise{In SAR, users normally observe a real scene without through any devices;}{In SAR, users normally do not require devices to observe a scene;} thus, no optical focal modulation is available.
On the other hand, we can consider synthetic focal modulation, where the appearance of a blurred real scene is synthesized by convolution computation, as in VST-AR systems, and is reproduced by superimposed projected imagery.
Researchers have developed various radiometric compensation techniques that enable pixel-wise color manipulation of a textured real surface \cite{doi:10.1111/cgf.13387,huang2019end}, which could be used to reproduce the blurred appearance.
However, these techniques cannot make a real surface appear completely blurred due to the limited dynamic range of current projectors \cite{10.1111:j.1467-8659.2008.01175.x}. In addition, the current projectors project pixels that are larger than underlying surface details \cite{6756996,7265057}.

% 3
% この研究では、bokehglassesを提案する。光学的にボケをdepth-independentにコントロールする
%spatially-varyingly controlling the saliency of real world for
% correct for the focal lengths of human eyes to
In this paper, aiming to realize innovative vision augmentation experiences, we propose IlluminatedFocus, a technique that optically controls the perceived blurs of real-world appearances in a spatially varying manner regardless of the distance from the user's eyes to observed real objects.
%A user observes a real scene with defocus blur regardless of the distance between the glasses to the scene.
As an example of images perceived by a user, a part of a real object appears blurred, while fine details of the other part at the same distance remain visible.
The core contribution of this research is a new computational display framework\footnote{The computational display framework involves joint design of hardware and display optics with computational algorithms and perceptual considerations \cite{MASIA20131012}.} that enables the depth-independent blurring of real-world appearances at a high spatial resolution.
Specifically, we apply \revise{electrically focus tunable lenses}{ETL} as eyeglasses and a synchronized high-speed projector as illumination for a real scene.
We periodically modulate the focal lengths of the glasses (focal sweep) at greater than 60 Hz so that a wearer cannot perceive the modulation.
A part of the scene that should appear focused is illuminated by the projector when it is in the focal range of the user's eyes, while another part that should appear blurred is illuminated when it is out of the focus.
Figure \ref{fig:teaser}(b) shows an experimental proof of the proposed technique.

As the basis of our spatial focus control, we construct a mathematical model to compute the degree of defocus (i.e., the size of the blur circle) of a real object in the proposed system.
Based on the model, we derive the blur range of a user wearing ETLs of a given optical power. Given the blur range, we can compute an appropriate optical power to switch the appearance of a real object between focused and blurred.
Based on the blur range, we discuss a design guideline to determine the range of the focal sweep.
We also model the apparent size of a real scene that is altered by the focal length modulation.
This leads to an undesirable visible seam between focused and blurred areas.
We solve this unique problem by gradually blending the two areas.
The blending weights are determined to diminish the seam with the smallest blending region.
Finally, we implement vision augmentation applications to demonstrate the feasibility of our proposal.

To summarize, the primary contributions of this paper are as follows.

\begin{itemize}
  \item We introduce the IlluminatedFocus technique, an innovative computational display framework that achieves depth-independent spatial defocusing of real-world appearances via focal sweep eyeglasses and a high-speed projector.
  \item We construct a mathematical model to compute the blur range of a user's eye to establish a design guideline to determine effective illumination timings and the range of the focal sweep.
  \item We propose a blending technique to diminish a conspicuous seam between blurred and focused areas.
  \item We confirm the feasibility of the IlluminatedFocus technique by implementing vision augmentation applications, such as visual guidance and concealing undesired visual information.
\end{itemize}

\section{Related Work}

% 1章で書いたように、ボケを使うといろいろなインタラクションが可能
% ARでも実装された例がある
%visual guidance \cite{doi:10.1080/17470218.2012.722659,Hata:2016:VGU:2909132.2909254,mti3010019}, focus and context (F+C) visualization \cite{963286,974515}, concealing undesired visual information (e.g., privacy protection) \cite{Yao:2013:FMA:2491367.2491377,10.1007/978-3-319-25554-5_25,8014910}, and enhancing the depth perception of a displayed image \cite{4480749,okatani,10.1007/978-3-642-23834-5_1}.

%Synthetic blur has been employed in a wide range of GUI applications as discussed in Section 1.
%Recognizing that it is also essential for AR applications, researchers have developed various interactive AR systems based on the synthetic blur.
Recognizing that synthetic blur is essential for AR applications, researchers have developed various interactive AR systems based on depth-independent spatial focus control of real-world appearances.
Specifically, these AR systems fall into the following three categories: (1) visual guide, (2) F+C visualization, and (3) diminished reality (DR).

% これらは、一般的なAR(矢印とか)でgaze/attention directionするものに比べて、重畳情報に注目点が隠されることがないことや、context情報が得られるという利点がある
(1) Typical AR systems employ graphical widgets such as virtual arrows, to draw the user's attention to real-world objects \cite{Biocca:2006:AFO:1124772.1124939,5336486,1544664,RUSCH2013127}.
These systems potentially destroy the visual experience by superimposing distracting overlays on the real scene.
On the other hand, HCI researchers have proved that a subtle image modulation can also effectively direct the user's gaze, such as luminance and color modulation \cite{Bailey:2009:SGD:1559755.1559757} and synthetic blur \cite{Veas:2011:DAI:1978942.1979158}.
McNamara proposed applying a synthetic blur technique to a mobile AR system to direct the user's gaze to specific areas of a real scene by blurring out unimportant areas \cite{McNamara:2011:EAH:2087756.2087853}.
This approach has an advantage of drawing a user's attention without significantly interrupting the visual experience.
(2) F+C visualization allows a user to focus on a relevant subset of the data while retaining the context of surrounding elements.
%enables applications to naturally draw the attention of a user to objects in the focus while still perceiving surrounding contextual information.
%Previous HCI researches showed that synthetic blur is one of the suitable techniques to realize this concept \cite{963286,974515,doi:10.1111/cgf.12877}.
Kalkofen et al. proposed an interactive F+C visualization framework for AR applications \cite{4538846,4569839}.
Their proposed framework applies a blur effect to suppress visual clutter in context areas and successfully supports a user to comprehend the spatial relationships between virtual and real-world objects.
(3) Hiding real-world objects is useful in various AR application scenarios, and researchers have worked extensively on DR techniques \cite{1240705,5643572,5643590,Iwai2011,Iwai:2006:LDS:1180495.1180519}.
%The most relevant application so far is privacy protection where personal information such as the face of a person, the license plate of a car, and the appearance in a private room is visually concealed from captured images \cite{Yao:2013:FMA:2491367.2491377,10.1007/978-3-319-25554-5_25,8014910}.
%To realize the concept, researchers have intensively worked on diminished reality (DR) techniques \cite{1240705,5643572,5643590,Iwai2011}.
Typical DR methods replace or fill in an undesired object (e.g., AR markers \cite{5643590}) with its background texture to make the object invisible. %completely removes the presence of the object.
%On the other hand, there are some cases where it is better to keep the presence of a diminished object than completely remove it.
%For example, when walking around a site where undesirable real objects are diminished, a user may hit these invisible objects.
%Another example scenario is a face-to-face meeting where a private paper document, which must be protected from a part of meeting members, is on a meeting table.
% DRは完全に存在自体を消してしまう。それはマーカー等を消す分には良いだろう。しかし実際に使うことを考えると、ある程度そこに消した対象の存在感は残したほうがよいと考えられる（ぶつかったりするので）
% preserve real world context
A blur effect has also been applied as a DR technique.
This approach preserves the presence of a diminished object compared to typical DR techniques.
In their X-ray vision AR system, Hayashi et al. proposed protecting the privacy of a person by blurring their face and body \cite{HAYASHI2010125}.

In all the above-mentioned previous systems, blurred real-world appearances are displayed on VST displays, such as HMDs and smartphones.
%All the previous AR systems employed VST-AR for applying a blur to a real-world appearance.
Because the blur effect can be implemented by a simple video signal processing, depth-independent, and spatially varying blur can be synthesized and displayed on VST displays.
On the other hand, as discussed in Section 1, there is no simple solution to realize such a flexible focus control in any vision augmentation platforms including OST-AR and SAR.
Because users see the real world directly in these systems, a blur needs to be controlled optically.
Previous studies have proposed optical solutions for blurring a real object in a spatially varying manner.
Himeno et al. placed a lens array plate in front of an object and spatially switched the state of the lens array to a flat transparent plate by filling transparent liquid having the same refractive index as that of the lens array \cite{Himeno:2018:FPS:3279778.3279784}.
Blur Mirror is a media art installation consisting of a motorized mirror array \cite{blur_mirror}.
An object reflected by the mirrors is selectively blurred by spatially varying the vibration of the mirrors.
Although these systems succeeded in spatially blur a real object, a relatively large optical setup needs to be positioned between the user and the object.
When either the blurring target or the user moves, they need to physically readjust the setup to keep their spatial relationship consistent.
In this paper, we relax this physical constraint by applying wearable glasses and a computational illumination, by which we can continuously provide a desired blur effect even when the object or user moves.

We apply ETLs to the eyeglasses used in the proposed system.
Previous research integrated such focus tunable lenses in AR and VR displays  \cite{7014259,Konrad:2017:ACN:3072959.3073594,8456852,Chang:2018:TMD:3272127.3275015,4637321,Jo:2019:TPL:3355089.3356577}.
The majority of these studies used the lenses to solve the vergence-accommodation conflict of typical HMDs.
Among them, the work most related to our research realized multifocal displays by driving the lenses to sweep a range of focal lengths at a high frequency and switching displayed content of different focal planes in synchronization with the focal sweep using a high-speed display \cite{8456852,Chang:2018:TMD:3272127.3275015,4637321,Jo:2019:TPL:3355089.3356577}.
The proposed system consists of devices similar to those used in previous studies and also applies fast focal sweep.
However, we use a high-speed projector to illuminate real objects to control the blur intensities of their appearances in a spatially varying manner rather than displaying virtual objects at different focal planes.
%it illuminates real objects by a high-speed projector such that areas of the objects to appear focused (blurred) are illuminated when the focusing distances of a user's eyes correspond (do not correspond) to them.
\revise{}{We demonstrated the first prototype at an academic conference \cite{Ueda:2019:IVA:3355049.3360530}. In the current paper, we describe the technical details and evaluate the proposed method qualitatively and quantitatively.}

\section{Depth-Independent Spatial Focus Control of Real-World Appearances}
\label{sec:method}

\begin{figure}[t]
  \centering
  \includegraphics[width=0.98\hsize]{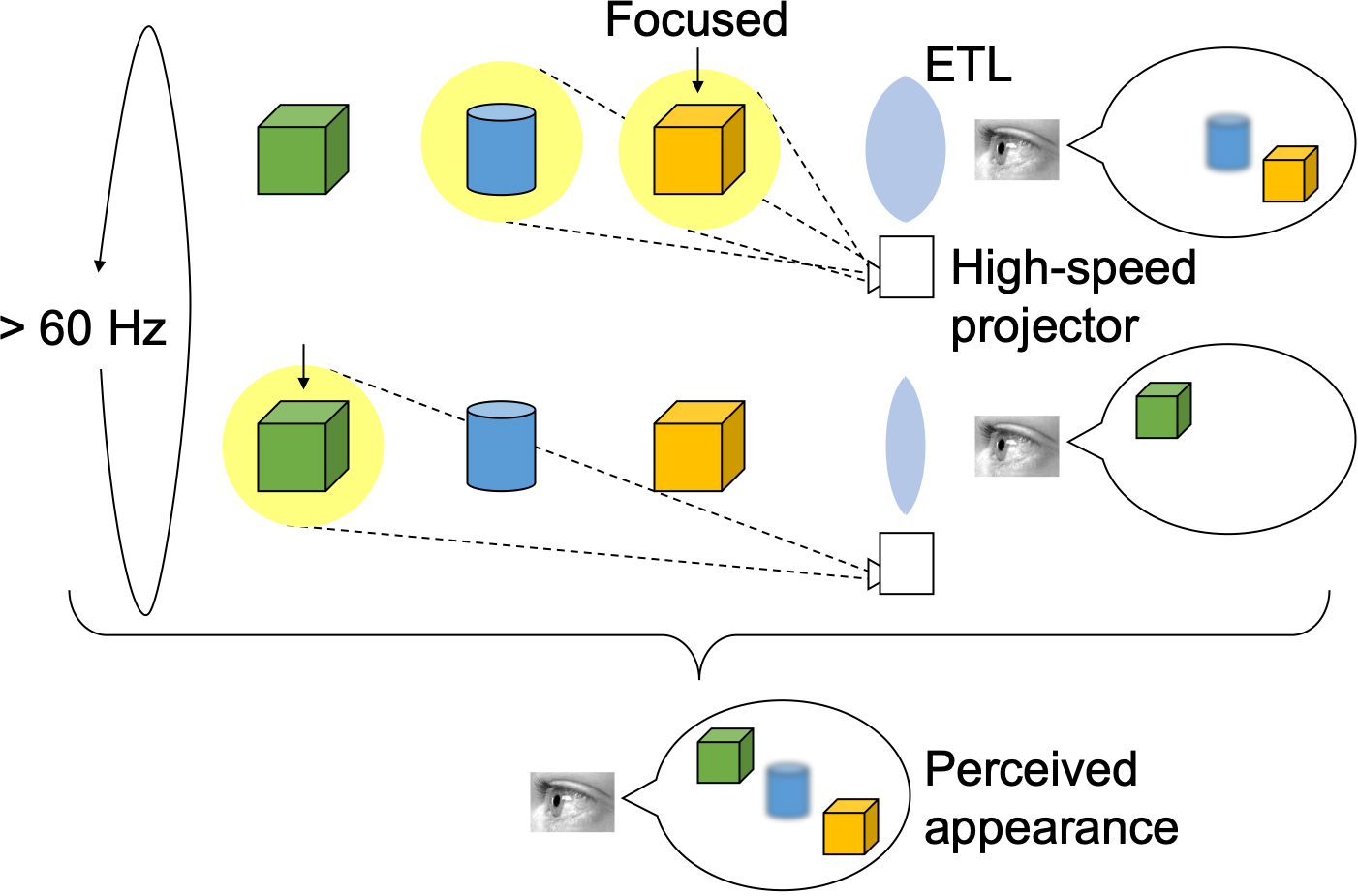}
  \caption{Principle of the proposed technique. \revise{}{The ETL focal length is modulated periodically at 60 Hz. The blue cylinder is illuminated when it is defocused, while the green and orange boxes are illuminated when they are focused. Consequently, an observer perceives an image in which only the blue object placed in the middle appears blurred.}}
  \label{fig:principle}
\end{figure}

% ここのfigureは、いつも使っているボーリングピンとかのやつを出す
We realize depth-independent spatial focus control by the fast focal sweep of human eyes and synchronized high-speed illumination (Figure \ref{fig:principle}).
Here, focal sweep refers to an optical technique that periodically modulates the optical power of an optical system (observer's eyes in our case) such that every part of an observed real scene is focused once in each sweep.
%Assuming a dark room environment, a part of the scene to appear focused is illuminated by the synchronized illumination for a short period when it is in focus of the observer's eyes, while another to appear blurred is illuminated when it is out of the focus.
If part of the scene is illuminated by a synchronized illuminator only when it is in the focal range of the observer's eyes, it appears focused.
On the other hand, if another part is illuminated only when it is out of focus, it appears blurred.
When the periodic optical power modulation is performed at higher than the critical fusion frequency (CFF), the observer does not perceive the modulation and the blink of the illumination.
We apply ETLs to periodically modulate the optical power of observer's eyes.
Note that here we assume that the observer is wearing ETL glasses.
A high-speed projector is used as the illuminator, which leads to the focus control of the observer's eyes on a per-pixel basis; thus, at a high spatial resolution.

In the rest of this section, we model the image formation of real-world objects in a user's eye with ETLs, as a mathematical basis of our technique (Section \ref{sec:method:model}).
Then, we discuss the blur range of the user with a given optical power of the ETL (Section \ref{sec:method:range}).
When real objects are within this range, they become blurred on the retina of the user.
Based on the blur range, we describe a design guideline for effective illumination timings and the range of the optical power modulation in the focal sweep (Section \ref{sec:method:guideline}).
Finally, we discuss a method to alleviate visible seams between the focused and blurred areas (Section \ref{sec:method:scaling}).
Note that, in this paper for simplicity and without loss of generality, we consider only the positive optical powers of the ETL.
The methods can be directly extended to negative optical powers.

%The resolution of the focal length modulation is determined by the focal sweep frequency of the ETL $r_E$ and the frame rate of the projector $r_p$.
%First of all, $r_E$ must be equal or higher than CFF $r_C$, thus $r_E\geq r_C$.
%Assuming that the focal length is linearly modulated (i.e., the temporal modulation pattern is a triangular wave) and $r_p$ is a constant multiple of $r_E$, the sweep range of the focal length is equally divided into $N_f$ ($=r_p/r_E$) parts by the projected illumination.
%Therefore, to achieve a fine level of focus control, the frame rate $r_p$ needs to be as high as possible.
%For example, assuming that the focal sweep is performed at 60 Hz, we need the frame rate of 600 FPS for the projector to provide ten focal ranges (i.e., $r_p=600$, s.t. $r_E=60$ and $N_f=10$).
%In the following, we describe a mathematical model of image formation in observer's eyes by assuming that the resolution of focal length modulation is sufficiently high.

\subsection{Image formation of real-world objects in the human eye with ETL}
\label{sec:method:model}

As a mathematical basis of our technique, we model the image formation of real world objects in a human eye with ETLs.
Specifically, the model computes the size of the blur circle of a point in a real scene on the retina of a user's eye.
The human eye consists of several refracting bodies such as the cornea, aqueous humor, vitreous humor, and crystalline lens.
We consider these together as a single lens and the retina as an image plane without loss of generality \cite{8458263}.
When a point in a real scene is defocused, it is imaged as a spot (i.e., a blur circle) on the image plane called a circle of confusion (CoC).

\begin{figure}[t]
  \centering
  \includegraphics[width=0.98\hsize]{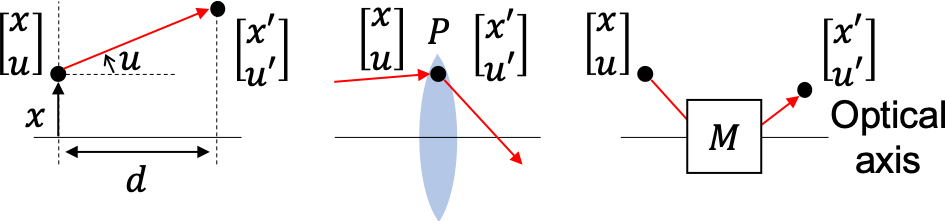}
  \caption{RTM analysis. (left) A ray \revise{transfers}{passes through} space. \revise{}{Note that $u'=u$.} (middle) A ray \revise{transfers}{passes through} a thin lens. (right) Two points are conjugate to each other for a given RTM.}
  \label{fig:rtm}
\end{figure}

We compute the size of blur circle $D_r$ on the image plane (i.e., retina) based on ray transfer matrix (RTM) analysis (Figure \ref{fig:rtm}).
RTM analysis is a mathematical tool used to perform ray tracing calculations under paraxial approximation.
The calculation requires that all ray directions are at small angles $u$ relative to the optical axis of a system such that the approximation $\sin u\simeq u$ remains valid.
An RTM is represented as follows:
\begin{equation}
  \begin{bmatrix}
    x'\\
    u'
  \end{bmatrix}
  =M
  \begin{bmatrix}
    x\\
    u
  \end{bmatrix}
  =
  \begin{bmatrix}
    A & B\\
    C & D
  \end{bmatrix}
  \begin{bmatrix}
    x\\
    u
  \end{bmatrix}
  ,
\end{equation}
where $M$ is an RTM and a light ray enters an optical component of the system crossing its input plane at a distance $x$ from the optical axis and travels in a direction that makes an angle $u$ with the optical axis.
After propagation to the output plane that ray is found at a distance $x'$ from the optical axis and at an angle $u'$ with respect to it.
If there is free space between two optical components, the RTM is given as follows:
\begin{equation}
  T(d)=
  \begin{bmatrix}
    1 & d\\
    0 & 1
  \end{bmatrix}
  ,
\end{equation}
where $d$ is the distance along the optical axis between the two components.
Another simple example is that of a thin lens whose RTM is given by
\begin{equation}
  R(P)=
  \begin{bmatrix}
    1 & 0\\
    -P & 1
  \end{bmatrix}
  ,
\end{equation}
where $P$ is the optical power (inverse of focal length) of the lens.

\begin{figure}[t]
  \centering
  \includegraphics[width=0.85\hsize]{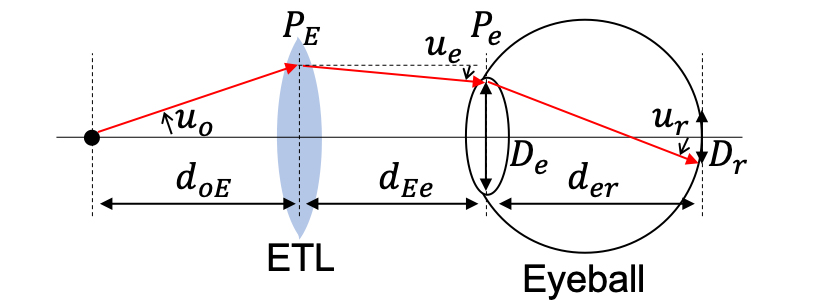}
  \caption{Parameters in the RTM analysis of the proposed system.}
  \label{fig:image_formation}
\end{figure}

% このfigureは今回の光学設定をのせる。目、ETL、対象（光軸上の点）、という感じ、式4と6の説明になる
A user of our system wears ETLs as eyeglasses such that the eye and the ETL share the same optical axis (Figure \ref{fig:image_formation}).
The size of the blur circle on the image plane is computed by tracing the marginal ray from a point on a real object.
Assuming that the ETL is larger than the pupil of the eye, the marginal ray passes at the edge of the pupil.
The angle of the marginal ray $u_o$ at the object can be computed by solving the following equation
\begin{equation}
  \begin{bmatrix}
    \frac{D_e}{2}\\
    u_e
  \end{bmatrix}
  =T(d_{Ee})R(P_E)T(d_{oE})
  \begin{bmatrix}
    0\\
    u_o
  \end{bmatrix}
  ,
\end{equation}
where $D_e$, $u_e$, $d_{Ee}$, $P_E$, and $d_{oE}$ represent the diameter of the pupil, the angle of the ray at the eye, the distance between the ETL and the eye, the optical power of the ETL, and the distance between the object point and the ETL, respectively.
Thus,
\begin{equation}\label{eq:u_o}
  u_o=\frac{D_e}{2(d_{oE}+d_{Ee}-d_{oE}d_{Ee}P_E)}.
\end{equation}
Finally, size of the blur circle on the image plane is computed by solving the following equation
\begin{equation}\label{eq:RTM_d_blur}
  \begin{bmatrix}
    \frac{D_r}{2}\\
    u_r
  \end{bmatrix}
  =T(d_{er})R(P_e)T(d_{Ee})R(P_E)T(d_{oE})
  \begin{bmatrix}
    0\\
    u_o
  \end{bmatrix}
  ,
\end{equation}
where $u_r$, $d_{er}$, and $P_e$ represent the angle of the ray at the image plane (i.e., retina), the distance between the lens of the eye and the retina, and the optical power of the lens of the eye, respectively.
Substituting Equation \ref{eq:u_o} in Equation \ref{eq:RTM_d_blur} gives the resultant size of the blur circle as follows:
\begin{equation}\label{eq:d_r}
  D_r=D_e\Bigl|1-d_{er}P_e+d_{er}\frac{1-d_{oE}P_E}{d_{oE}+d_{Ee}-d_{oE}d_{Ee}P_E}\Bigr|.
\end{equation}
%

% フォーカス領域
\subsection{Blur range of human eye with ETL}
\label{sec:method:range}

\begin{wrapfigure}{r}{4.5cm}
  \centering
  \includegraphics[width=0.98\hsize]{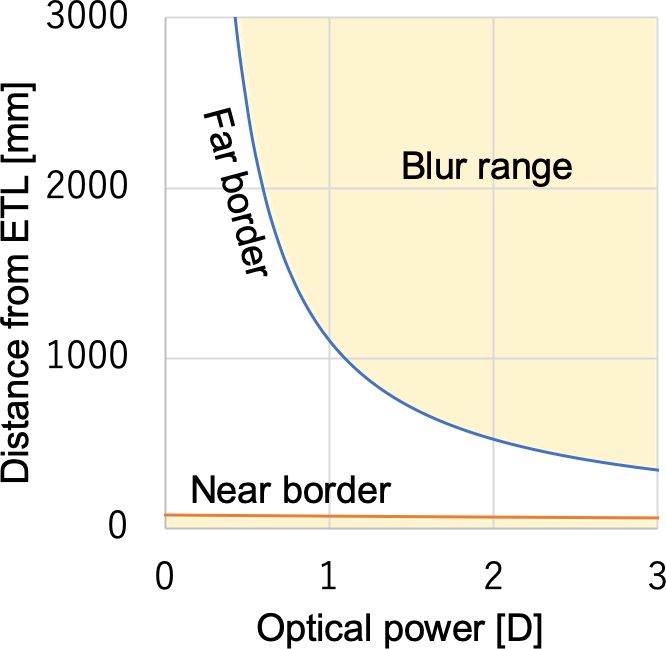}
  \caption{Blur range based on Equation \ref{eq:range}.}
  \label{fig:model_blurrange}
\end{wrapfigure}

For each optical power of the ETL $P_E$, we can compute the range of the distance from the ETL to a point on the optical axis, within which real objects become blurred on the retina.
The blur range is determined by two factors: accommodation and DOF.
Accommodation refers to the process by which the human eye changes its optical power to maintain focus on an object as its distance varies from the far point (the maximum distance) to the near point (the minimum distance).
We denote the optical powers for the far and near points as $P_e^f$ and $P_e^n$, respectively.
For each optical power of the eye, real objects are in acceptably sharp focus on the retina when they are within the DOF.
Suppose we denote the acceptable size of the CoC as $D_r^a$, the maximum and minimum distances of the DOF ($d_{oE}^{a+}$ and $d_{oE}^{a-}$, respectively) are computed using Equation \ref{eq:d_r} as follows:
\begin{equation}\label{eq:range}
  d_{oE}^{a\pm}=\dfrac{D_e(d_{Ee}d_{er}P_e-d_{Ee}-d_{er})\pm d_{Ee}D_r^a}{D_e(P_E(d_{Ee}d_{er}P_e-d_{Ee}-d_{er})-d_{er}P_e+1)\pm D_r^a(d_{Ee}P_E-1)}.
\end{equation}
When real objects are at points $d_{oE}^{a+}$ and $d_{oE}^{a-}$ distant from the ETL on the optical axis, the light from these objects hits the top or bottom endpoints of the acceptable CoC.
Then, the blur range for an optical power of the ETL $P_E$ is determined by two \revise{boarder}{border}s illustrated in Figure \ref{fig:model_blurrange} as:
\begin{itemize}
  \item $d_{oE}$ is less than $d_{oE}^{a-}$ subject to $P_e=P_e^n$ (near \revise{boarder}{border}).
  \item $d_{oE}$ is larger than $d_{oE}^{a+}$ subject to $P_e=P_e^f$ (far \revise{boarder}{border}).
\end{itemize}
%

% far borderとnear borderの理論値
Figure \ref{fig:model_blurrange} shows the blur range for different $P_E$ assuming the human eye as the reduced eye model \cite{reduced_eye}.
We found that the near \revise{boarder}{border} is less than 80 mm for all positive optical powers.
The proposed method requires a projector to illuminate real objects, and it is difficult for most commercially available projectors to provide such illumination when the objects are less than 80 mm away from a user's face.
Therefore, in the rest of this paper, we only consider the far \revise{boarder}{border} of the blur range.
Figure \ref{fig:model_blurrange} shows the $P_E$ value at which a real object at a certain distance from the ETL is blurred on the retina.
%In other words, we can compute a phase of the periodical focal sweep of the ETL, at which the projector should illuminate the object to make it appear blurred.

\subsection{Design guideline of illumination timing and focal range sweep}
\label{sec:method:guideline}

\begin{figure}[t]
  \centering
  \includegraphics[width=0.98\hsize]{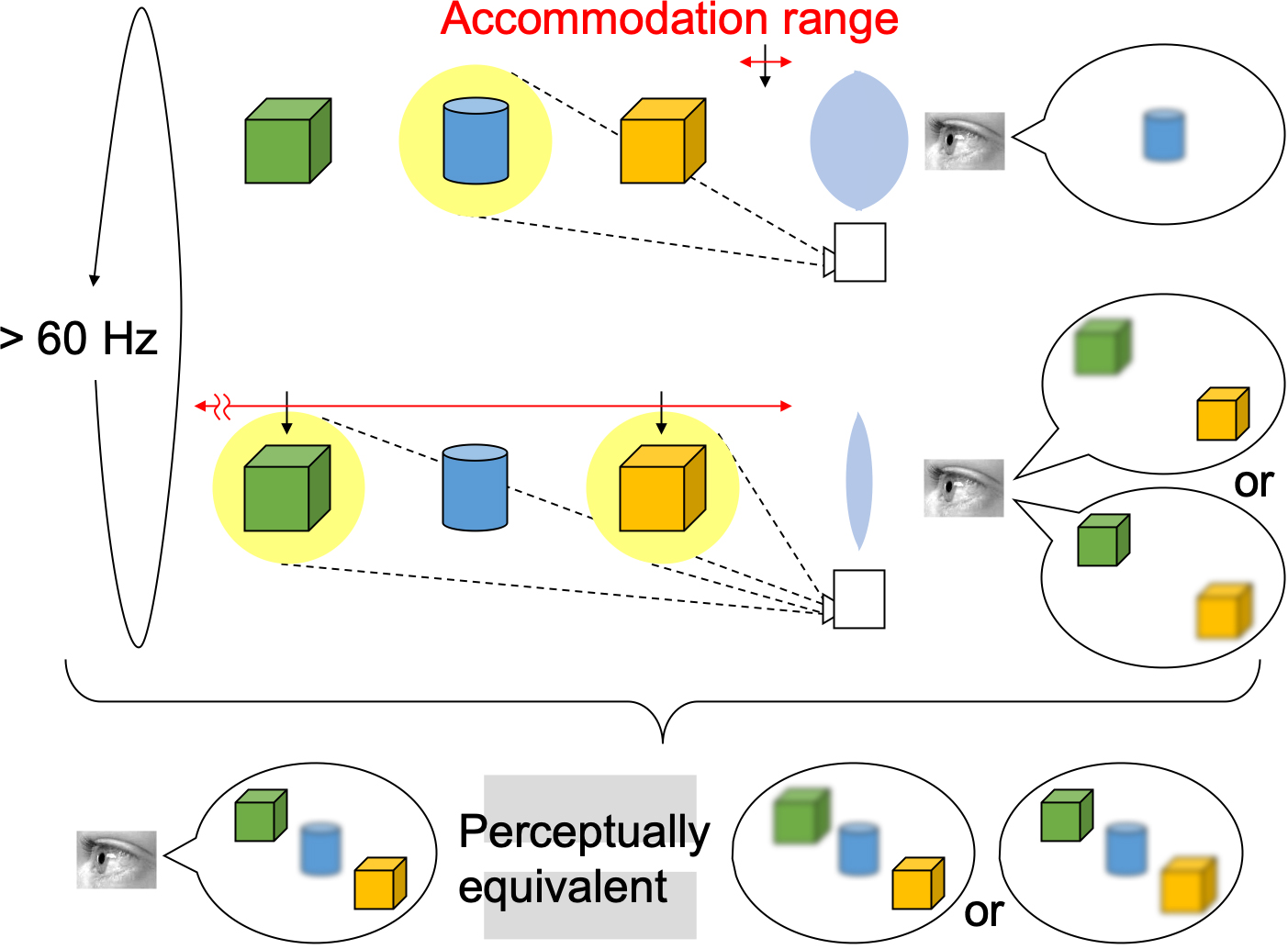}
  \caption{\revise{Effective illumination timings}{Illumination timings considering accommodation. The focal length of the ETL is modulated at 60 Hz. Objects to appear blurred are illuminated when the optical power of the ETL is the greatest in the modulation. The other objects to appear focused are simultaneously illuminated when the optical power is zero. These objects appear focused when the observer gazes at them. This is perceptually equal to a situation where these objects appear focused simultaneously}.}
  \label{fig:principle2}
\end{figure}

Figure \ref{fig:model_blurrange} provides an insight into human vision, i.e., a human with normal vision can accommodate an object located from near to far distances, when the optical power of the ETL is low.
This means that it is sufficient to illuminate objects to appear focused only when the optical power is zero (i.e., $P_E=0$).
In this case, when a user looks at an object (and thus, focuses on it), only those objects located at the same distance are in focus at the same time.
When the user looks at another object, the previously focused objects become out of focus.
However, in most cases, \revise{our conscious is only on an object which we look at}{our attention is only on an object that we are looking at}; thus, a situation where objects we look at are always in focus is perceptually equal to these objects being in focus simultaneously.
Figure \ref{fig:model_blurrange} provides another insight, i.e., an object placed at any distance can be blurred when the optical power $P_E$ is sufficiently large.
Therefore, as a guideline for illumination timing, we illuminate objects to appear focused when $P_E=0$ and those to appear blurred when $P_E>0$ (Figure \ref{fig:principle2}).

A large $P_E$ requires a wide focal sweep range.
This leads to a long period in the optical power modulation because a typical ETL physically changes the lens thickness to modulate its optical power.
Because the sweep frequency needs to be larger than the CFF in the proposed method, the sweep range cannot be increased excessively.
%In addition, as we discuss in the next section, a large $P_E$ leads to a perceivable change in the apparent size of the real object on the retina.
In addition, the wider the focal sweep range is, the more the optical power changes within one frame of the projector.
To illuminate real objects accurately at a desired optical power, the sweep range should not be increased.

Therefore, the focal sweep range needs to be both as small as possible and it is sufficiently large to make a target real object appear blurred.
Assume a typical situation where (1) several real objects are located at different positions in a real scene, (2) the distances from these objects to the ETL are known, and (3) some objects appear blurred.
We determine the sweep range in the following two steps.
First, we compute the minimum optical power value required to make each appear blurred.
This value is on the ``far \revise{boarder}{border}'' of Figure \ref{fig:model_blurrange}.
Then, we select the maximum optimal power among the minimum values.
Note that we refer to the selected optical power as $P_E^s$ in the rest of the paper.
We determine the focal sweep range as 0 diopter (D) to $P_E^s+\alpha$, where $\alpha$ is a user-defined offset.

% 像の拡大縮小
\subsection{Alleviating visible seam caused by apparent scaling of real-world objects}
\label{sec:method:scaling}

\begin{figure}[t]
  \centering
  \includegraphics[width=0.98\hsize]{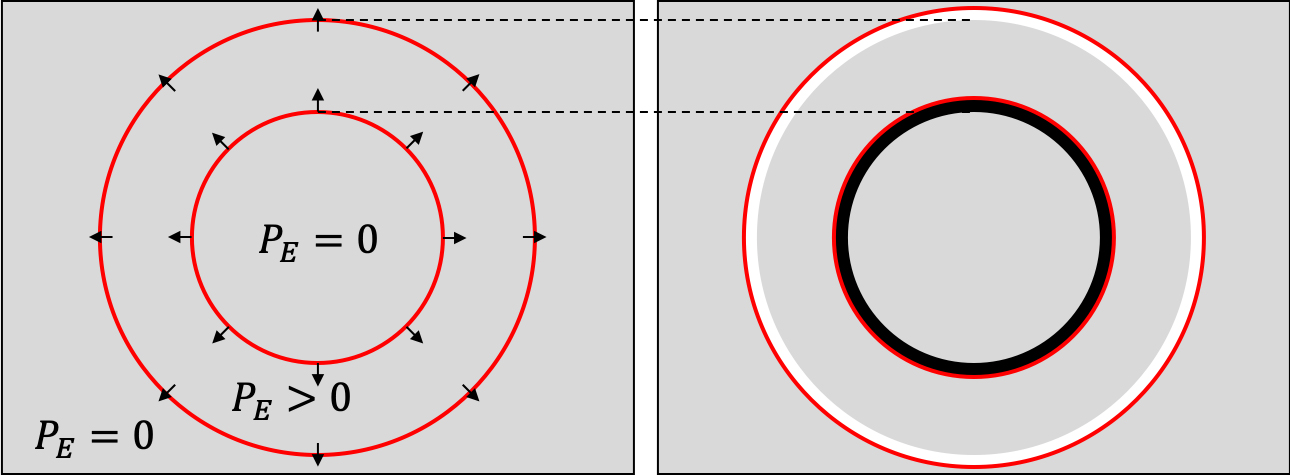}
  \caption{Visible seam by apparent scaling of a blurred area. Left: the doughnut region (red lines) is illuminated when $P_E>0$. Consequently, the doughnut region appears blurred and becomes larger relative to the optical center. Right: the apparent scaling causes the gap and overlap between the doughnut and the other regions, resulting in dark and bright seams, respectively.}
  \label{fig:seam}
\end{figure}

When we see a real-world object through typical eyeglasses that correct for either myopia or hyperopia, the apparent size of the object becomes smaller or larger.
The same phenomenon occurs in the proposed system where the ETLs are used as eyeglasses.
A unique problem with the proposed system is that the apparent scaling of a real scene is not spatially uniform.
More specifically, the apparent size of a real object differs spatially, when different optical powers $P_E$ are applied to different parts of the object.
For example, when the system makes only a certain area of the object appear blurred by illuminating this area when $P_E>0$ and the other area when $P_E=0$, only the blurred area becomes larger relative to the optical center.
When the area where $P_E>0$ is located closer to the optical center than the area where $P_E=0$, a gap occurs between these areas and appears as a dark seam (Figure \ref{fig:seam}).
On the other hand, when the two areas are in the reverse locations, these areas overlap and the overlapped area appear as a bright seam.
This section describes a method to alleviate the seams.

\begin{figure}[t]
  \centering
  \includegraphics[width=0.98\hsize]{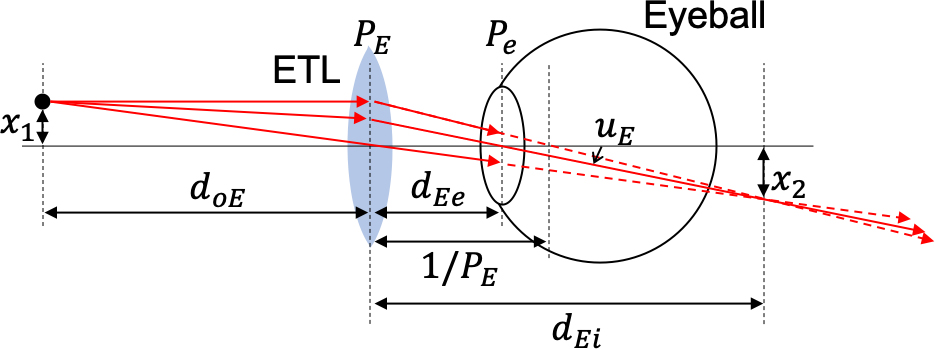}
  \caption{Ray trancing of the composition of the ETL and eye.}
  \label{fig:scaling}
\end{figure}

% まず、どれくらい大きくなる・小さくなるのか
First, we discuss the extent to which the apparent size of a real object is changed (i.e., scaling factor) by the ETL using a ray tracing technique.
% このfigureは、上田君のuist資料の複合レンズの図を使う
As shown in Figure \ref{fig:scaling}, light rays from a real object are refracted at the ETL such that they form a real image.
The human eye observes these refracted rays.
The distance from the ETL to the image $d_{Ei}$ is obtained by the thin lens formula:
\begin{equation}\label{eq:d_ei}
  \frac{1}{d_{oE}}+\frac{1}{d_{Ei}}=P_E\ \rightarrow\ d_{Ei} = \frac{d_{oE}}{d_{oE}P_E-1}.
\end{equation}
Suppose the height of the object from the optical axis is $x_1$, that of the image $x_2$ is obtained by the similarity of triangles and substituting Equation \ref{eq:d_ei} as follows:
\begin{equation}\label{eq:h_2}
  x_1:x_2 = d_{oE}:d_{Ei}\ \rightarrow\ x_2 = \frac{x_1}{d_{oE}P_E-1}.
\end{equation}
For the human eye, the visual angle of the real object is determined by a refracted ray passing through the center of the lens.
Thus, the visual angle $u_E$ is obtained by the following equation
\begin{equation}
  \tan u_E=\frac{x_2}{d_{Ei}-d_{Ee}}.
\end{equation}
Substituting Equations \ref{eq:d_ei} and \ref{eq:h_2}, the angle is computed as follows:
\begin{equation}
  u_E = \arctan\left(\frac{x_1}{d_{oE}+d_{Ee}-d_{oE}d_{Ee}P_E}\right).
\end{equation}
Therefore, the scaling factor $s$ of the real object under $P_E>0$ compared to the object under $P_E=0$ can be computed as follows:
\begin{equation}\label{eq:s}
  s=\frac{\tan u_E}{\tan u_{E=0}}=\frac{d_{oE}+d_{Ee}}{d_{oE}+d_{Ee}-d_{oE}d_{Ee}P_E}.
\end{equation}

Second, we describe the method to alleviate a seam caused by the apparent scaling of a part of a real object.
We apply a simple feathering or blending technique.
Suppose there are two areas next to each other, one without scaling (i.e., $P_E=0$) and the other with scaling (i.e., $P_E=p>0$) that are illuminated at $t_0$ and $t_p$, respectively.
We can calculate the seam region as the difference of the scaled area between before and after scaling using Equation \ref{eq:s}.
We illuminate the seam region using the high-speed projector both at $t_0$ and $t_p$.
At $t_0$, the intensity of the illumination in the seam region is decreased linearly from the unscaled area (intensity=1.0) to the scaled area (intensity=0.0).
%ブレンディングの説明（いらないかも）
At $t_p$, the intensity is decreased linearly from the scaled to the unscaled area to ensure that the sum of these contributions becomes 1.0 at any seam region.

\section{System Evaluation}
\label{sec:exp}

We evaluated the proposed technique using a prototype system.
First, we investigated how accurately our mathematical models predicted the size of a blur circle and the blur range (Sections \ref{sec:exp:size} and \ref{sec:exp:blurrange}, respectively).
Then, we determined if the proposed technique achieved spatial defocusing (Section \ref{sec:exp:spatialdef}).
Finally, we evaluated how well visible seams between blurred and focused regions could be alleviated (Section \ref{sec:exp:seam}).

\subsection{Experimental setup}
\label{sec:exp:setup}

\begin{figure}[t]
  \centering
  \includegraphics[width=0.98\hsize]{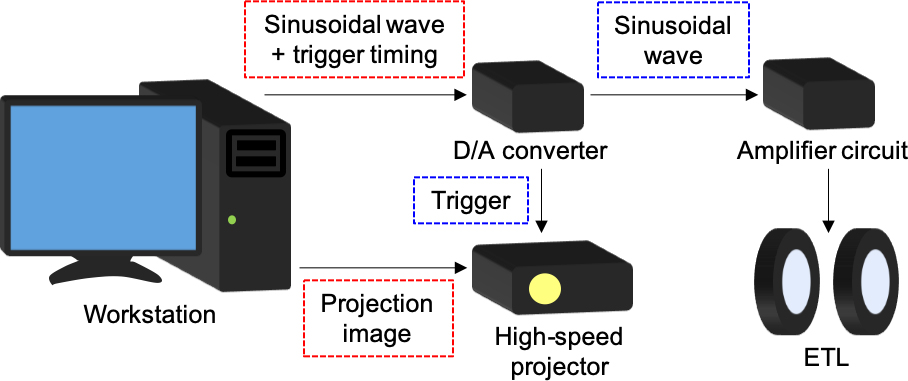}
  \caption{System configuration (red box: digital signal, blue box: analog signal).}
  \label{fig:system}
\end{figure}

We constructed a prototype system consisting of a pair of ETLs and a synchronized high-speed projector (Figure \ref{fig:teaser}(a)).
We used polymer-based liquid lenses as the ETLs.
The polymer-based ETLs achieve faster focal change than other types of ETL while maintaining a relatively large aperture size.
Consequently, they have been exploited in a wide range of optical systems, from micro-scale systems, such as microscopes, to larger-scale systems, such as HMDs \cite{Konrad:2017:ACN:3072959.3073594,8456852,Chang:2018:TMD:3272127.3275015,4637321} and projectors \cite{7014259}.
Specifically, we inserted two ETLs (16 mm aperture, Optotune AG, EL-16-40-TC) in an eyeglass frame fabricated from an FDM 3D printer to form a wearable (69$\times$128$\times$67 mm, 200 g) device (Figure \ref{fig:teaser}(a)).
The optical power of the ETL was controlled from -10 D to 10 D by changing the electrical current.
The digital signal generated by a workstation (CPU: Intel Xeon E3-1225 v5@3.30GHz, RAM: 32 GB) was input to a D/A converter (National Instruments, USB-6343) and converted to analog voltage.
This voltage was then converted to an \revise{elecrtric}{electric} current by a custom amplifier circuit using an op-amp (LM675T).
Finally, the current was fed to the ETL.
According to the ETL's data sheet, the input analog voltages in our system ranged from -0.07 to 0.07 V.
We employed a consumer-grade high-speed projector (Inrevium, TB-UK-DYNAFLASH, 1024$\times$768 pixels, 330 ANSI lumen) that can project 8-bit grayscale images at 1,000 frames per second.
Projection images were generated by the workstation and sent to the projector via a PCI Express interface.
The display timing of each projection image was adjusted by a 5 V trigger signal from the workstation via the D/A converter.
\revise{}{The system configuration is depicted in Figure \ref{fig:system}. We assume that our system works in a dark environment.}

\begin{figure}[t]
  \centering
  \includegraphics[width=0.98\hsize]{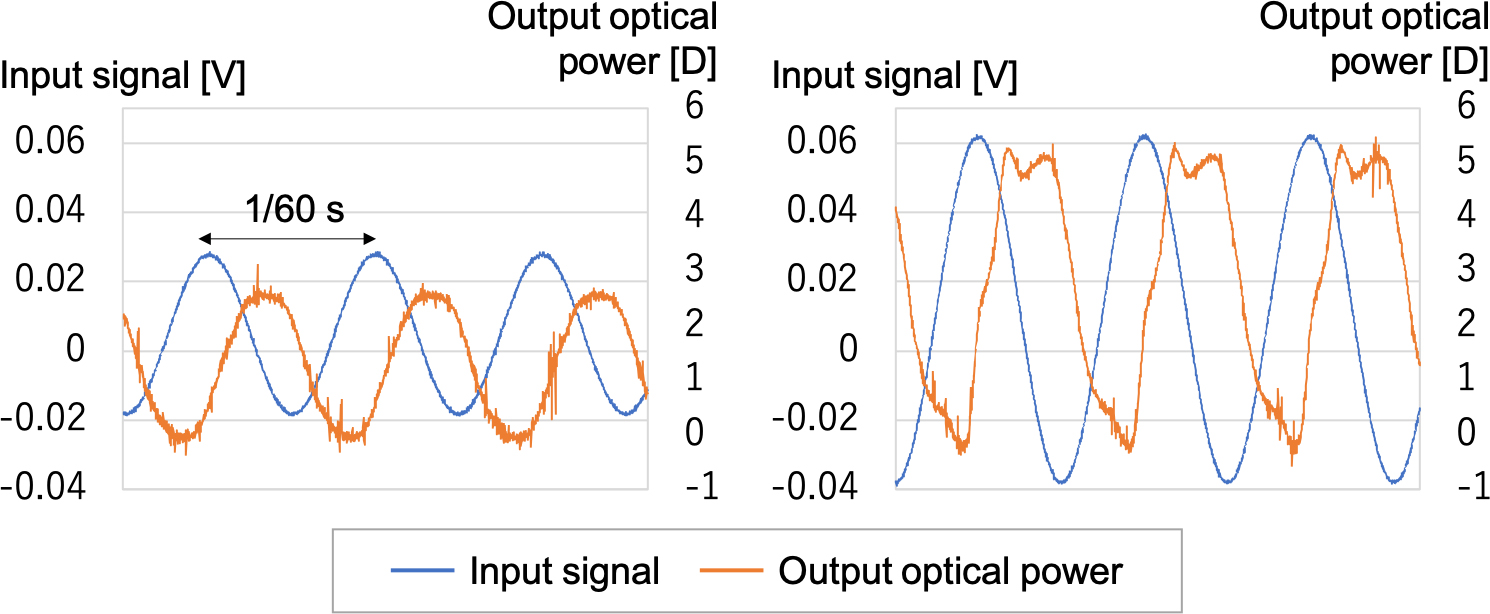}
  \caption{Measured optical powers by input waves with different ranges.}
  \label{fig:wave}
\end{figure}

The IlluminatedFocus system performed a periodic focal sweep by applying a sinusoidal wave as an input signal to the ETLs.
The frequency of the wave was set to 60 Hz throughout the experiment.
We applied waves of different offsets and amplitude to the ETL and measured the resulting optical powers.
We prepared 71 input voltage values (from -0.07 to 0.07 V at 0.002 V intervals) and used every combination of these values (2485 in total) as the maximum and minimum values of the input sinusoidal waves.
The optical power was measured using a photodiode and a laser emitter (see our previous paper \cite{izawa} for more details).
%The optical power was measured using a photodiode placed about 5 mm behind the ETL \cite{izawa}.
%A laser emitter (wavelength: 635 nm, output power: 0.9 mW) was placed in front of the ETL.
%The laser beam passed through the ETL and intercepted the photodiode.
%Raising the optical power of the ETL reduced the laser spot size on the photodiode (i.e., increased the laser power density), enhancing the electric current created in the photodiode, from which we indirectly measured the optical power.
% 計測した波のグラフをいくつか用意する。(複数波長分）
Figure \ref{fig:wave} shows two examples of the time series of the measured optical powers by input waves with different ranges.
As shown in this figure, we found that the output values were periodical; however, they did not form clean sinusoidal waves.
% 同じ波形のグラフから、１周期分だけ取り出したような画像にする。
Therefore, we stored one period of each output wave along with the corresponding input wave in a database in order to look up the optical power at a given phase of the input wave.
% 計測波のジオプトリレンジをグラフで一括表示したい
%The range of the output optical power varies according to the range of the input wave (Figure xxx).
%Generally, the wider the range of the optical power is, the more the optical power changes within 1 frame of the projector (i.e., 1 ms).
%To illuminate real objects accurately at a desired optical power, the range should be as small as possible.
Once a target range of optical power is given, we determined the input wave by searching the database for one having the narrowest range among those capable of generating the target range of the optical power.

To synchronize the ETLs and the high-speed projector, we used the same photodiode to measure the delay of the high-speed projector from a trigger signal of the workstation to the actual projection.
As a result, we found that the delay was 0.46 ms.
Using this delay information and the data in the database, we can use to projector to illuminate a real object exactly when the ETLs' optical powers are the target optical power.
%When the ETLs are driven by a sinusoidal wave, we can illuminate a real object by the projector exactly when the ETLs' optical powers are a target one by their optical powers at each phase obtained from the database and the delay information of the projector.
\revise{We conducted a preliminary test by ourselves and determined}{We conducted a preliminary test and determined} 0.2 D as the offset value $\alpha$ (Section \ref{sec:method:guideline}).

\subsection{Blur circle size}
\label{sec:exp:size}

% いくつかの距離おいた対象が、それぞれの焦点位置で、どのような見えになるのかを確認して、レンズモデル通りかどうかと、きちんとボケるかどうか（つまり同期がとれてるかどうか）を確認する。

\begin{figure}[t]
  \centering
  \includegraphics[width=0.98\hsize]{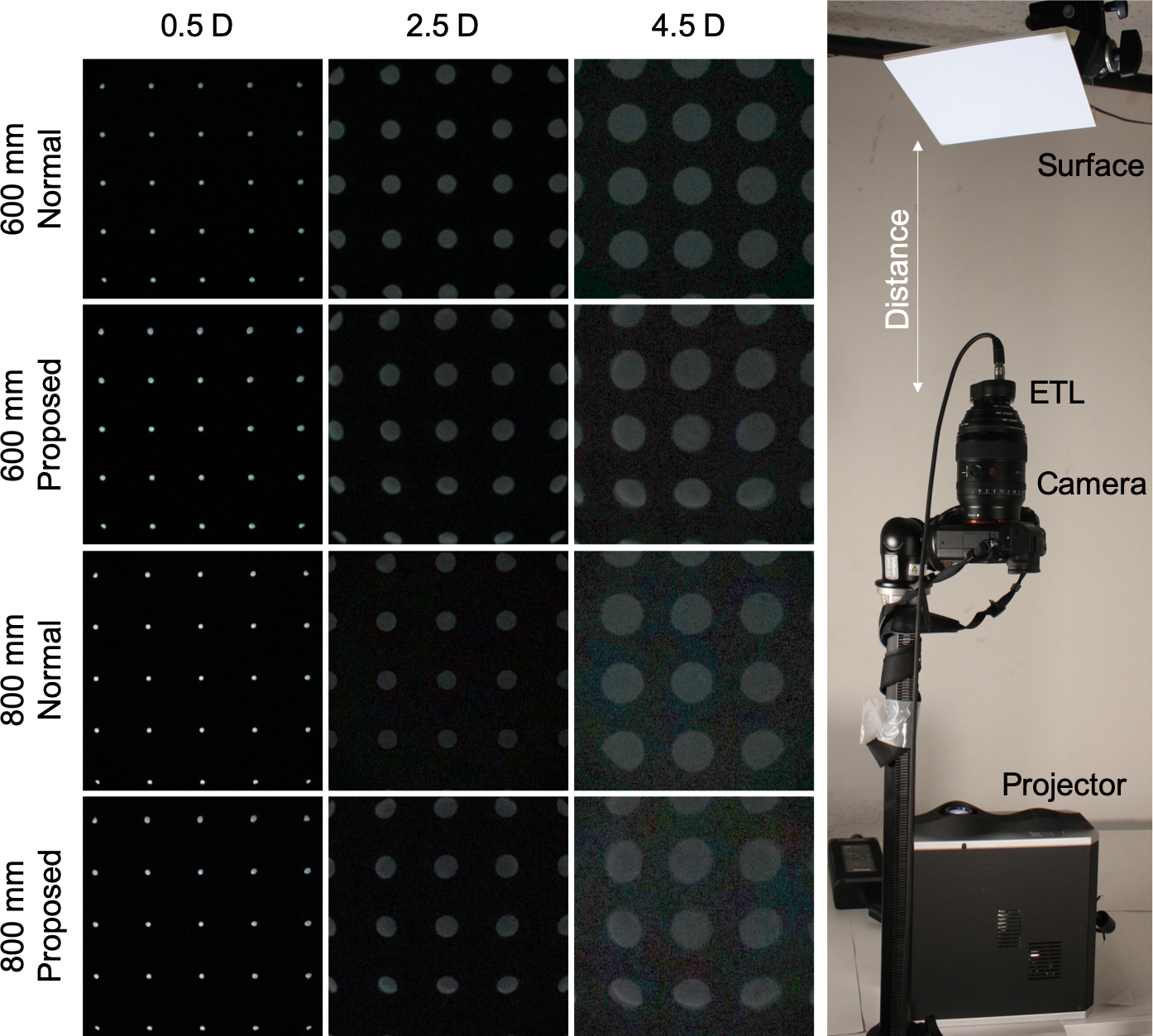}\\\ \\
  \includegraphics[width=0.98\hsize]{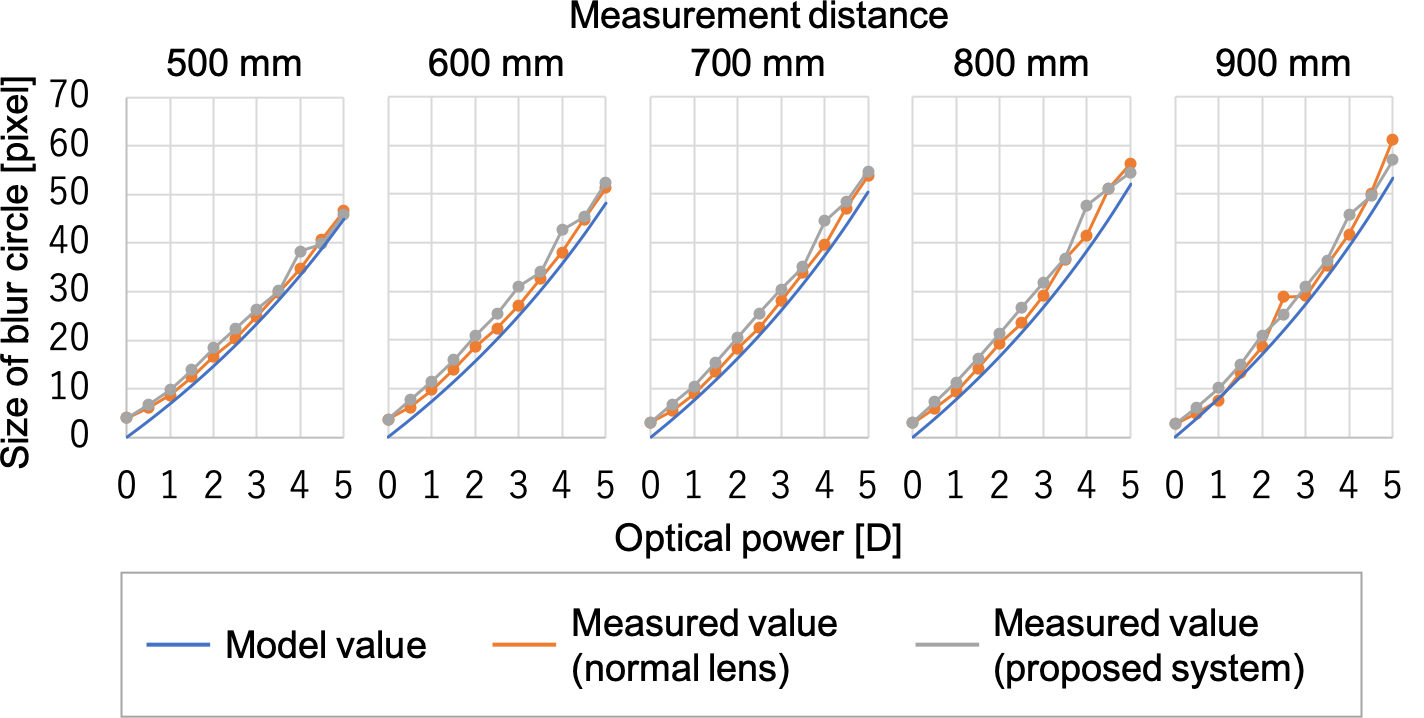}
  \caption{Blur size measurement. Top left: captured dot pattern on the surface (contrast and brightness adjusted). Top right: measurement setup. Bottom: measured radius of the blur circle. Note that the brightness of the dot pattern images are adjusted for better visibility.}
  \label{fig:res_blursize}
\end{figure}

We evaluated how a real scene appeared blurred using the proposed system.
% whether or not presented images by our system follow the image formation model in Section \ref{sec:method:model}.
We measured a dot pattern on a planar surface using a camera (Sony $\alpha$7S II, lens: Sony FE 24mm F1.4 GM) on which the ETL is mounted (Figure \ref{fig:res_blursize}).
%The dot was illuminated by the projector.
We prepared five measurement distances between the surface and the camera ($500,600,\ldots,900$ mm).
We measured the dot pattern with eleven different optical powers ($P_E=0.0,0.5,1.0,\ldots,5.0$) under two conditions: normal and proposed.
The normal condition was used as the baseline.
In the normal condition the dot pattern was observed with fixed ETL optical power.
In a preliminary test, we printed a black dot pattern on a sheet of white paper and observed it with different optical powers.
However, due to the low contrast of the printed media, the blur circles of the dots were not observable at more than 1.0 D.
Therefore, we  used the high-speed projector to project a dot pattern onto a uniformly white surface in a dark room.
The camera's exposure time was set to 1/60 s and the dot pattern was projected for 1 ms (i.e., one frame of the projector), and the optical power of the ETL was fixed.
In the proposed condition, we applied the focal sweep to the ETL and projected the same dot pattern for 1 ms when the optical power was one of the eleven $P_E$ values.
The input wave to the ETL was determined as discussed in Section \ref{sec:exp:setup}.
More specifically, we prepared ten input waves to generate target ranges of the output optical power by combining 0.0 D and the eleven optical powers (i.e., 0.0 to 0.0 D, 0.0 to 0.5 D, 0.0 to 1.0 D, ..., 0.0 to 5.0 D).
Note that the optical power of the ETL was actually not swept in the first target range (0.0 to 0.0 D).
The camera exposure time was the same as the normal condition.
% we applied eight different input waves for each of the target optical powers of $P_E=0.5,1.0,\ldots,4.0$.

Figure \ref{fig:res_blursize} shows the measured dot patterns.
The appearance of the real scene (the projected dot pattern) is similar under the normal and proposed conditions.
This result indicates that the proposed system successfully synchronized the ETL and the projector achieved focus control of real-world appearances that was the same as that of a normal lens.
More quantitatively, we \revise{}{binarized each captured image using Otsu's method \cite{4310076}, }picked the center $3\times3$ blur circles\revise{ from each captured image}{}, fit circles to the selected blur circles, and measured their radii.
Figure \ref{fig:res_blursize} shows the measured radii under both normal and proposed conditions with blur sizes computed using our image formation model (Equation \ref{eq:d_r}).
In this result, the measured radii were slightly larger but very close to the model values.
\revise{The slight offset caused due to that the projected dots were not infinitesimal points.}{The slight offset was due to the projected dots not being infinitesimal points.}
Taking this into account, we consider that our image formation model well predicted the size of the blur circle in the proposed system.
%, which was, on the other hand, assumed in our image formation model.
We computed the difference between the size of blur circles under normal and proposed conditions.
The average difference was 1.4 pixels with a standard deviation (SD) of 1.8 pixels.
This difference \revise{can be explained by that the optical power}{occurred because the optical power} of the ETL changed within a single frame of the dot pattern projection (i.e., 1 ms); consequently, the blur circles of different sizes were accumulated \revise{in a capturing process}{in the capture process}.

\subsection{Blur range}
\label{sec:exp:blurrange}

We measured the blur range of users of our system based on a typical visual acuity test.
When a participant could not distinguish the correct direction of a Landolt ring of the visual angle of one minute, we considered that the ring was located within the blur range of the participant's eye.
We printed an array of six Landolt rings in randomized directions on a piece of paper and placed it in front of the ETL.
We prepared different sizes of the rings according to the distance from the ETL to maintain the visual angles of the rings as one minute.
Each participant viewed the rings using their dominant eye through the ETL and reported the direction of the rings.
If more than three directions were correct, we considered that the current distance was not within the blur range.
The Landolt rings were placed at nine distances from the ETL ($500,750,\ldots,2500$ mm).
At less than 500 mm, the projected image region was too small to illuminate the Landolt rings in our setup.
However, 2500 mm was a sufficiently far distance for our intended applications (Section \ref{sec:app}).
For each distance, we changed the optical power of the ETL from 0 to 2.4 D at 0.2 D intervals (i.e., 13 optical powers).
Then, we recorded the maximum optical power at which each participant could distinguish the directions of more than three Landolt rings.
%Note that the optical powers were determined through a preliminary test.

\begin{figure}[t]
  \centering
  \includegraphics[width=0.98\hsize]{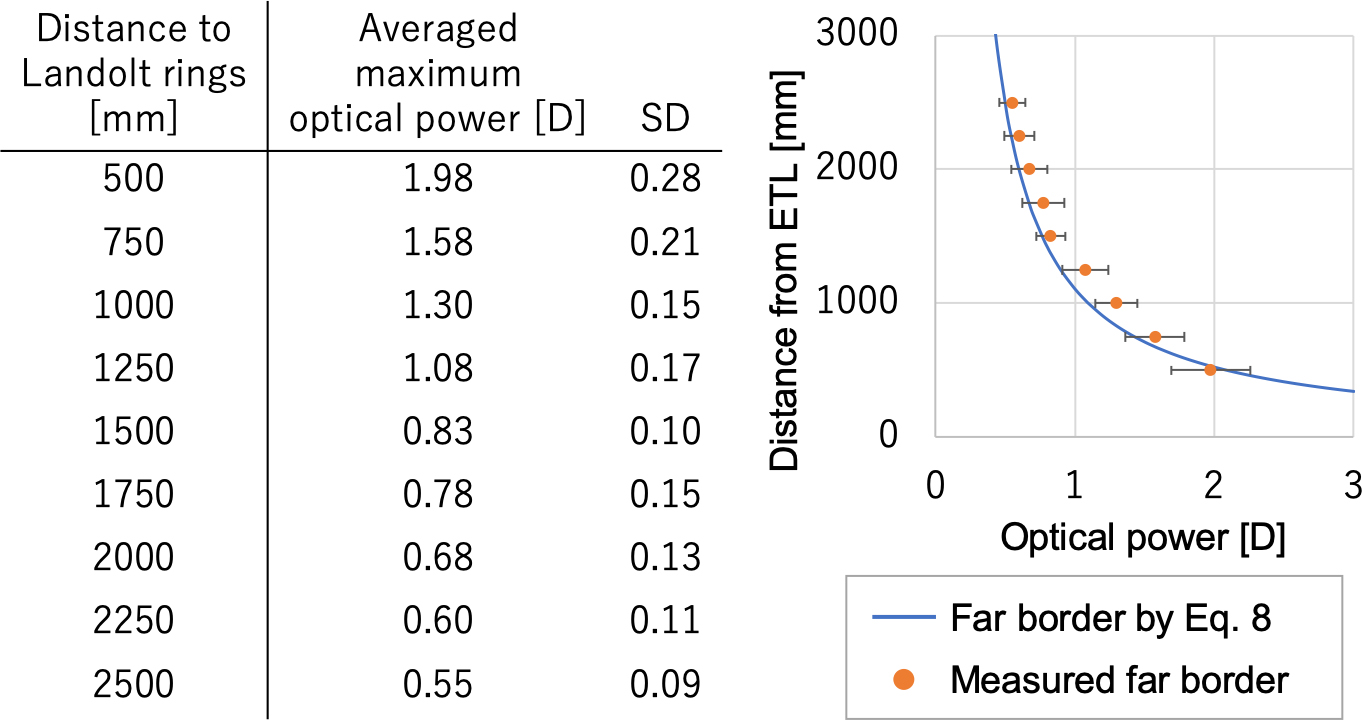}
  \caption{Blur range measurement. Left: averages and SD of maximum optical powers at which each participant could distinguish the directions of more than three (out of six) Landolt rings. Right: plotted data with the computational model of ``far \revise{boarder}{border}'' by Equation \ref{eq:range} (Figure \ref{fig:model_blurrange}).}
  \label{fig:res_blurrange}
\end{figure}

Eight participants were recruited from the local university (male: 7, female: 1, age: 22-32).
Six participants were nearsighted.
Their vision was corrected with the ETL by offsetting its optical power.
The optical power values in the following results are adjusted to consider the offset.
\revise{}{Throughout the experiment, each participant's head was fixed using a chin rest.}
Figure \ref{fig:res_blurrange}(left) shows the average and SD of the maximum optical power values at each distance.
We plotted the average values with the model ``far \revise{boarder}{border}'' (Figure \ref{fig:model_blurrange}) in Figure \ref{fig:res_blurrange}(right).
%We visualized the inverse function of the model of Equation \ref{eq:range} in the same figure.
This result indicates that the blur range can be accurately predicted by the proposed model (Equation \ref{eq:range}).
Therefore, in the subsequent experiments and applications, we used the model and the ``far \revise{boarder}{border}'' to compute the focal sweep range (Section \ref{sec:method:guideline}).

\subsection{Spatial defocusing}
\label{sec:exp:spatialdef}

\begin{figure}[t]
  \centering
  \includegraphics[width=0.98\hsize]{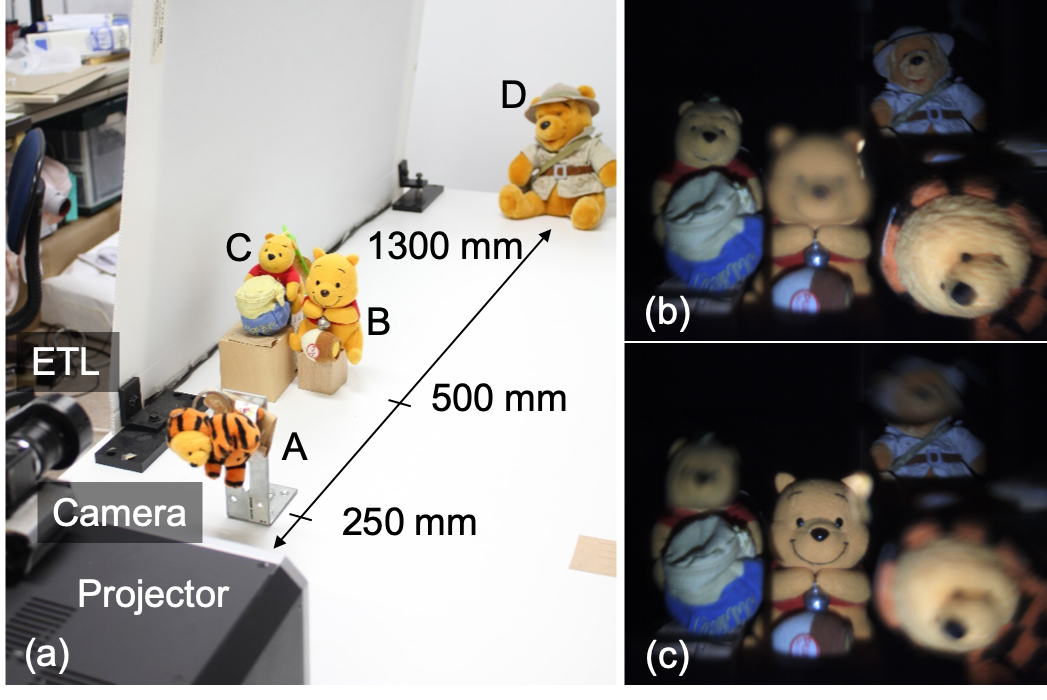}
  \caption{Evaluation of spatial defocusing: (a) Experimental setup. Captured results where (b) only \revise{the center object indicated by the red arrow}{object B} appears blurred and (c) only \revise{the center object}{object B} appears focused.}
  \label{fig:res_pooh}
\end{figure}

%プーさん
We investigated whether or not depth-independent spatial blur control is possible in the proposed system.
We placed four objects denoted A, B, C, and D at different locations as shown in Figure \ref{fig:res_pooh}(a).
The objects were 250, 500, 500, and 1300 mm, respectively, from the ETLs.
According to the design guideline in Section \ref{sec:method:guideline}, the focal sweep range was determined as 0 D to $P_E^s+\alpha$.
Then, objects to appear focused were illuminated when the optical power of the ETLs was 0 D, and objects to appear blurred were illuminated when it was $P_E^s+\alpha$.

We prepared two spatial defocusing conditions.
%In the first condition, we made only the object A appear blurred and the other objects appear focused.
%$P_E^s$ in this condition was the optical power on the ``far border'' of 500 mm (see Figure \ref{fig:model_blurrange}).
%In the second condition, we made all the objects appear focused; and thus, $P_E^s$ was 0 D.
In the first condition, only the center object appeared blurred and the other objects appeared focused.
$P_E^s$ in this condition was the optical power on the ``far \revise{boarder}{border}'' of 500 mm (Figure \ref{fig:model_blurrange}).
In the second condition, only the center object appeared focused and the other objects appeared blurred.
$P_E^s$ in this condition was the optical power on the ``far \revise{boarder}{border}'' of 250 mm.
We asked ten local participants to observe the objects through the ETLs under the two conditions.
\revise{}{\revise{}{As before, participant's heads were fixed during the experiment.}
After observation of each condition, we asked the participants to identify \revise{if they observed the intended appearance described above.}{which object appeared blurred.}
All the participants \revise{agreed it and none of them reported that they perceived \revise{flickers}{flickering}.}{responded that only object B appeared blurred under the first condition and objects A, C, and D appeared blurred under the second condition.}}

Figure \ref{fig:res_pooh} shows captured appearances of the objects under the two experimental conditions using a camera (Ximea MQ013CG-ON) attached to one of the ETLs.
To reproduce the perceived appearances, the illumination timings were different from the above.
Specifically, the projector illuminated each object to appear focused when it is in focus of the camera with the ETL, and another to appear blurred when it is out of focus.
For example, in the first condition, the projector illuminated objects A, B, C, and D when the focusing distances of the camera were 250, 500, 500, and 1300 mm, respectively.
%The focal sweep range was adjusted because the optical configuration of the camera is different from our human eye model.
All participants agreed that the captured image showed similar appearances they observed in the above mentioned user study.
Therefore, we confirmed that the proposed system achieved depth-independent spatial blur control.
In particular, it is optically impossible to produce the appearances of Figure \ref{fig:res_pooh} with normal lens systems.
These results verify the effectiveness of the proposed spatial defocusing technique.

\subsection{Alleviating visible seam}
\label{sec:exp:seam}

We conducted an experiment to investigate the effectiveness of our method to alleviate the visible seam caused by the apparent scaling of observed real objects.
We prepared eight experimental conditions (= 2 textures$\times$2 seams$\times$2 optical powers).
Specifically, we used two textured surfaces (document and picture) as observed objects.
Each paper was placed 500 mm away from the ETLs.
% two scaling patterns
As discussed in Section \ref{sec:method:scaling}, there are two types of seams (gap and overlap) according to the spatial relationship of the blurred and focused areas.
Therefore, two seam conditions were prepared for each texture.
% two optical powers (1D and 2D)
The width of a seam varies according to the optical power of the blurred area.
To check if the method works for different seam widths, we prepared two optical powers (1 D and 2 D) for the experiment.
For each condition, we compared the appearance of the surface with our method to that without our method.

The same participants (Section \ref{sec:exp:spatialdef}) observed the textured surfaces under the eight conditions.
\revise{}{\revise{}{As before, throughout the experiment, each participants' head was fixed using a chin rest.}
After observing a pair of appearances with and without our method in each condition, the participants were asked if the seam was alleviated by our method.
All the participants \revise{agreed it for all the conditions}{answered that the proposed method could alleviate the seam in all conditions}.}

\begin{figure}[t]
  \centering
  \includegraphics[width=0.98\hsize]{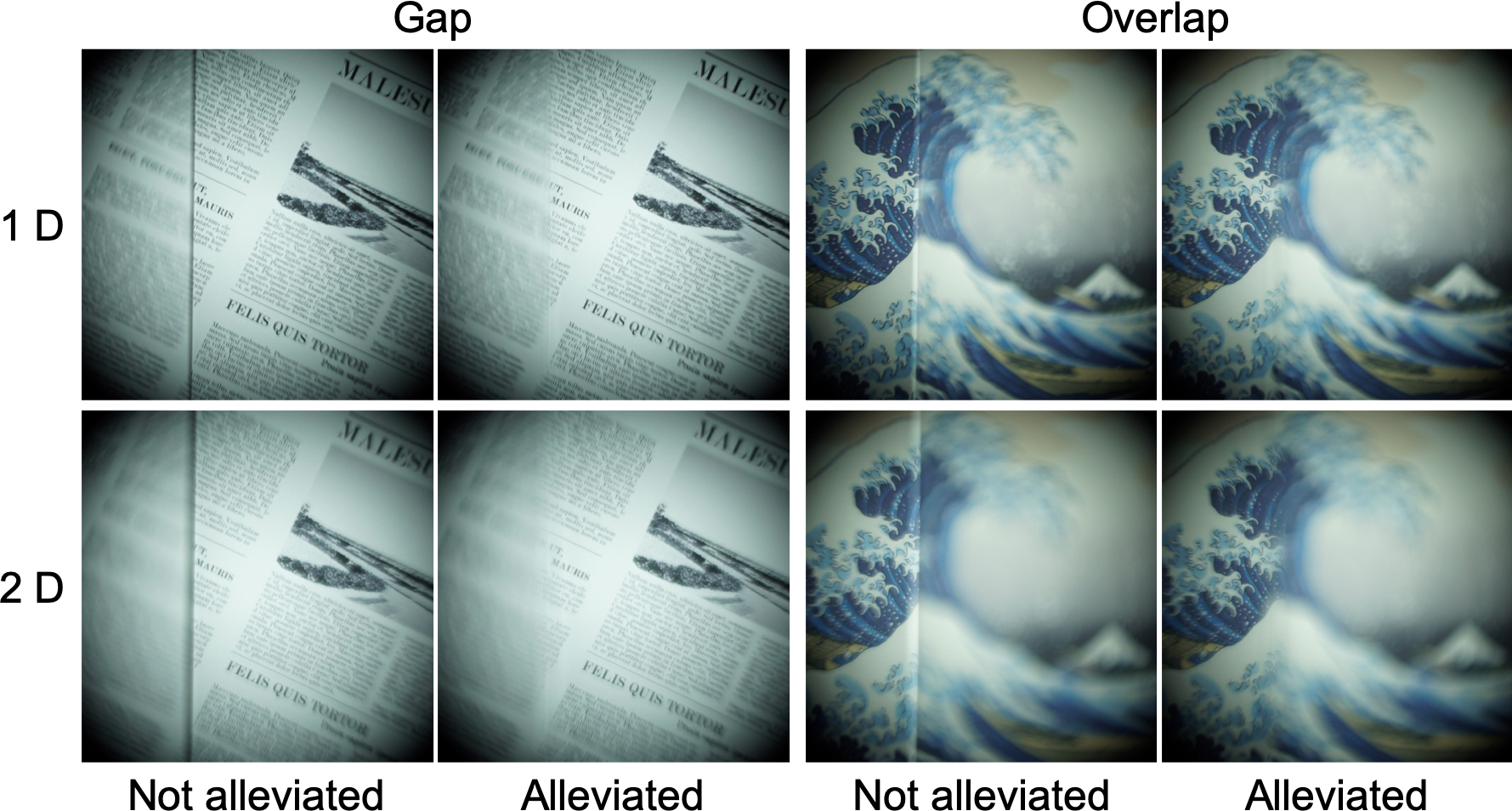}
  \caption{Captured results of \revise{blurred and focused areas next to each other}{the visible seam alleviation}.}
  \label{fig:res_seam}
\end{figure}

Figure \ref{fig:res_seam} shows captured appearances of the surfaces under all eight conditions using the same camera used in the previous experiment (Section \ref{sec:exp:spatialdef}).
All participants agreed that the figure shows appearances similar to what they observed.
From the captured results, we confirmed that the proposed method could successfully alleviate the visible seam caused by the apparent scaling of observed real objects.

\section{Applications}
\label{sec:app}

We developed four different vision augmentation application prototypes using the proposed IlluminatedFocus technique.
In this section, we show how they worked.
Please see the supplementary video for more details.
Note that we used an ETL-mounted camera with a lens adapter for the recording.
The adapter added a dark ring in the video images.
In actual applications, a user does not see the ring.

\subsection{Visual guide}

% visual guide
% トリケラトプス
% 楽譜ガイド
% 平木システム
\begin{figure}[t]
  \centering
  \includegraphics[width=0.98\hsize]{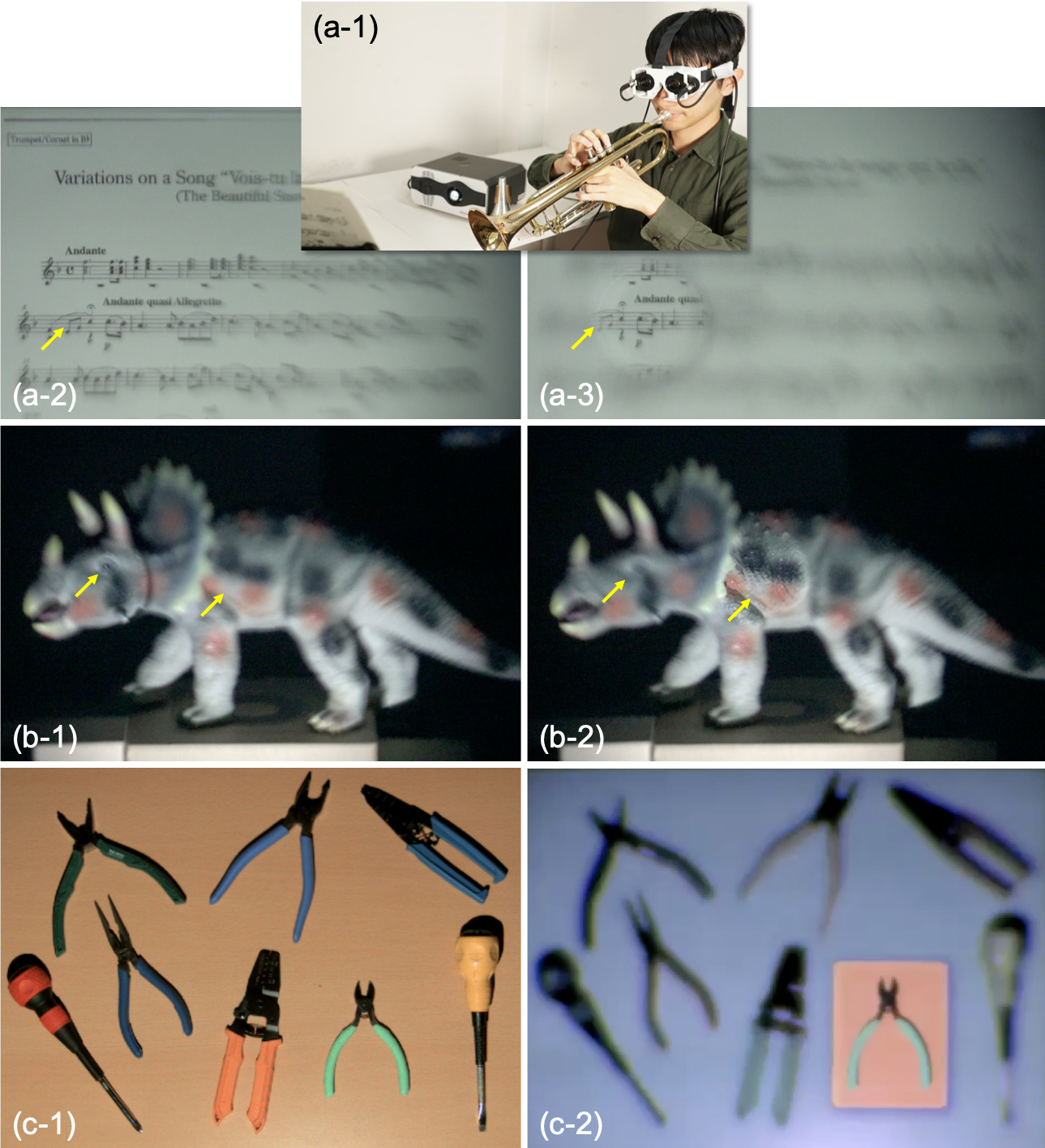}
  \caption{Visual guide applications: \revise{}{(a) Part of a musical score sheet appears in focus to support a practice session.} (b) Museum guidance where a curator explains the object by moving the focused area from its face to its body. (c) Tool selection support drawing a user's attention to \revise{the correct tool for a specific task}{a tool} by decreasing the saliency of other tools (i.e., blur and desaturation).}
  \label{fig:res_visguide}
\end{figure}

As the first application, we implemented three visual guidance systems that naturally direct the user's gaze to a specific part of a real object by making that part appear focused and the other parts appear blurred.
Figure \ref{fig:res_visguide}(a) shows the first example where a part of a musical score to be played is made to appear focused.
\revise{}{We assume that the musical score is scanned in advance and the notes on the score are recognized.}
The focused area automatically moves over the notes so that a player can use the proposed system to practice the score while keeping a desired tempo.
Figure \ref{fig:res_visguide}(b) shows the second example.
We assume a situation where a curator in a museum or a teacher in a class sequentially explains several parts of a target object to visitors or students.
The curator (teacher) moves the focused area \revise{}{manually using a pointing device (e.g., a touch panel)} according to the explanation to draw the visitors' (students') attention to this specific area.
Figure \ref{fig:res_visguide}(c) shows the last example.
In this example, a tool to be used in the next operation is made to appear focused.
We assume a situation where an inexperienced person uses the proposed system to assemble a complicated electrical system.
The person is not familiar with the tools and has no idea which one to use in each step.
Our system can support such a situation to draw the person's attention to the right tool.
In this application, we used a full-color high-speed projector (Texas Instruments, DLP LightCrafter 4500) and applied a radiometric compensation technique \cite{10.1111:j.1467-8659.2008.01175.x} to make the right tool appear focused and to decrease the color saturation of the other tools.
This example verifies the effectiveness of the combination of the proposed and SAR techniques.
\revise{From these systems of the visual guidance application, we confirmed that an advantage of our visual guidance applications is that they can draw the viewer's attention without adding any distracting graphical widgets such as virtual arrows.}{ Overlaying a virtual arrow or characters is an alternative solution for visual guidance. The visibility of the overlaid information depends on the appearance of the background. In case of a cluttered background, it is difficult for an observer to understand the information \cite{6381407}. In such situations, we believe that our blur-based approach provides better visual guidance.}

\begin{figure}[t]
  \centering
  \includegraphics[width=0.98\hsize]{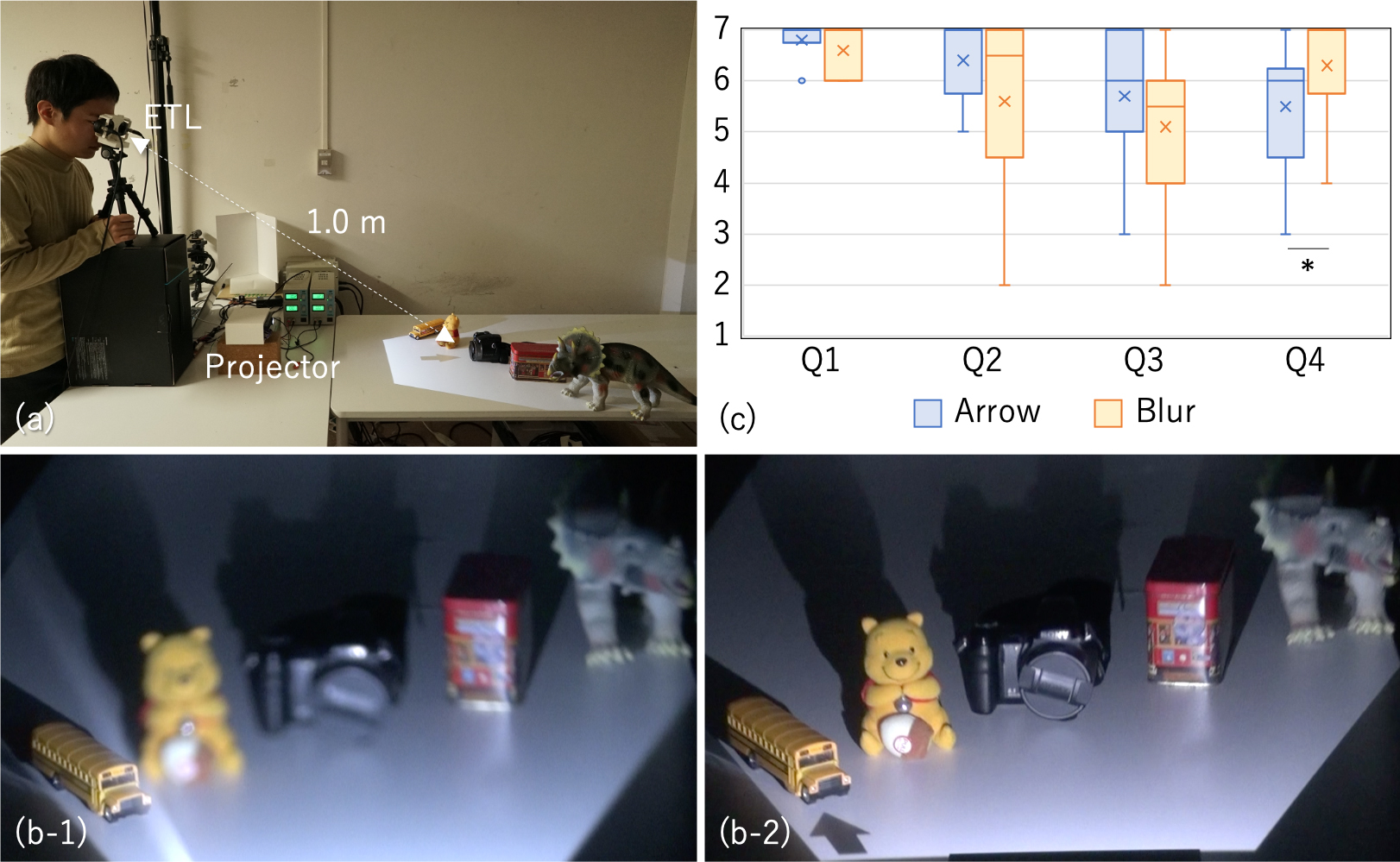}
  \caption{\revise{}{User study: (a) Experimental setup. (b) The appearances of the objects under the blur (b-1) and arrow (b-2) conditions, respectively. The leftmost object was indicated. (c) The box plots of the participants' answers ($*: p<0.05$).}}
  \label{fig:res_study}
\end{figure}

\revise{}{We conducted a user study to investigate how users react to our artificial blur in the visual guide application. We placed five objects on a table (Figure \ref{fig:res_study}(a)) and asked participants to gaze at an object which is indicated either by a projected arrow or spatial defocusing (i.e., only the indicated object appears focused). Thus, there were two experimental conditions regarding the indication method (i.e., {\it arrow condition} and {\it blur condition}). The object to be gazed was randomly switched at 2 seconds intervals. Each participant performed this task for 1 minute in each condition. Right after each task, the participant answered the following four questions based on 7-point Likert scale (1 = strongly yes, 7 = not at all):}\par
\revise{}{Q1: Did you notice flickers?}\par
\revise{}{Q2: Do you feel motion sickness?}\par
\revise{}{Q3: Are you tired?}\par
\revise{}{Q4: Were the indications difficult to understand?}\par
\noindent
\revise{}{The environment light was turned off in the experiment.}

\revise{}{Ten participants were recruited from a local university. They saw the objects through the ETLs which were fixed as shown in Figure \ref{fig:res_study}(a). The appearances of the objects in the both experimental conditions and the results were shown in Figure \ref{fig:res_study}(b-1, b-2, c). For each question, we performed a paired $t$-test between the result in the arrow condition and that in the blur condition. We confirmed that there was a significant difference only in the fourth question ($p<0.05$). From the results of Q1, Q2 and Q3, we confirm that the levels of the negative reactions to our artificial blur are as low as those to the normal projection mapping. In addition, the result of Q4 shows that a visual guide interface based on our artificial blur provides a better understandability to a user than the conventional GUI-based interface.}

\subsection{F+C visualization}

\begin{figure}[t]
  \centering
  \includegraphics[width=0.98\hsize]{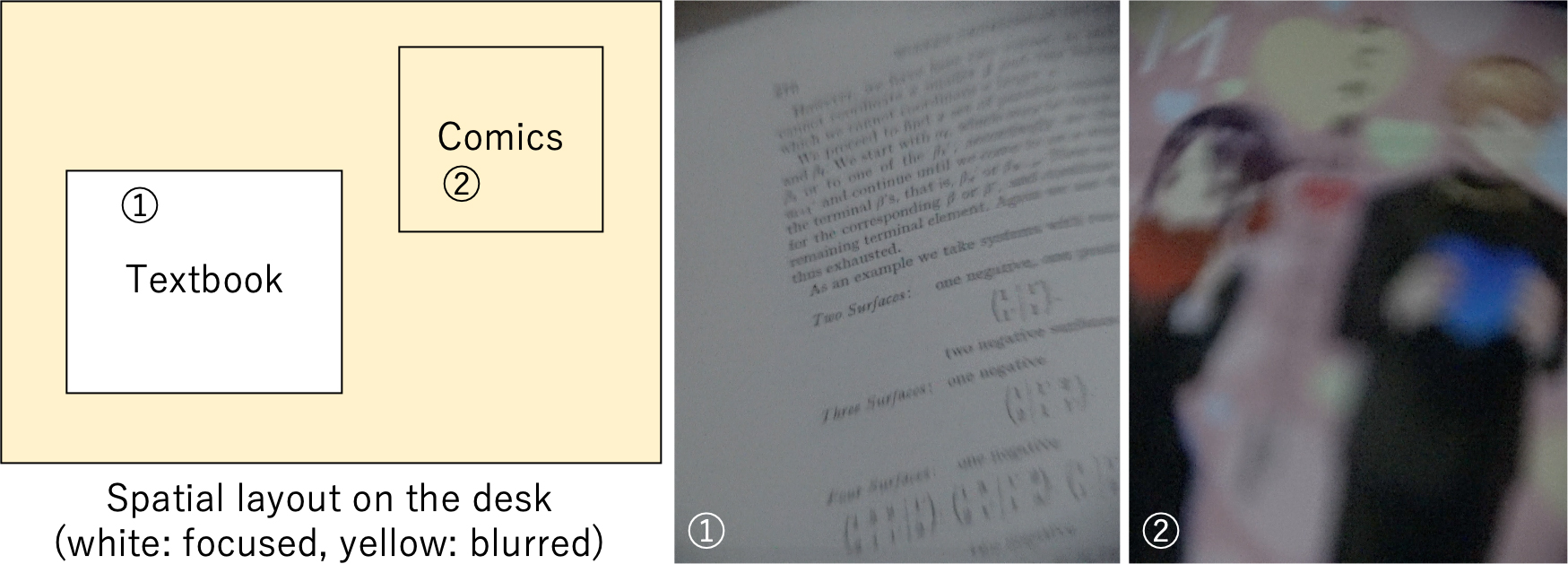}
  \caption{F+C visualization application. See supplementary video for better understanding of the spatial layout on the desk.}
  \label{fig:res_fpc}
\end{figure}

% F+C visualization
% 勉強集中
As the second application, we implemented an F+C visualization system.
This system supports a student studying at a desk to concentrate on reading textbooks and writing in notebooks.
The system makes the textbooks and notebooks appear focused and the other areas on the desk appear blurred.
Figure \ref{fig:res_fpc} shows some appearances of the desk in the system.
From the result, we confirmed that the system successfully suppressed visual clutters on the desk, and consequently provided a student with an effective learning environment where they could focus only on their studies.

\subsection{Diminished reality}

% concealing undesired visual information / DR
% シワ消し
\begin{figure}[t]
  \centering
  \includegraphics[width=0.98\hsize]{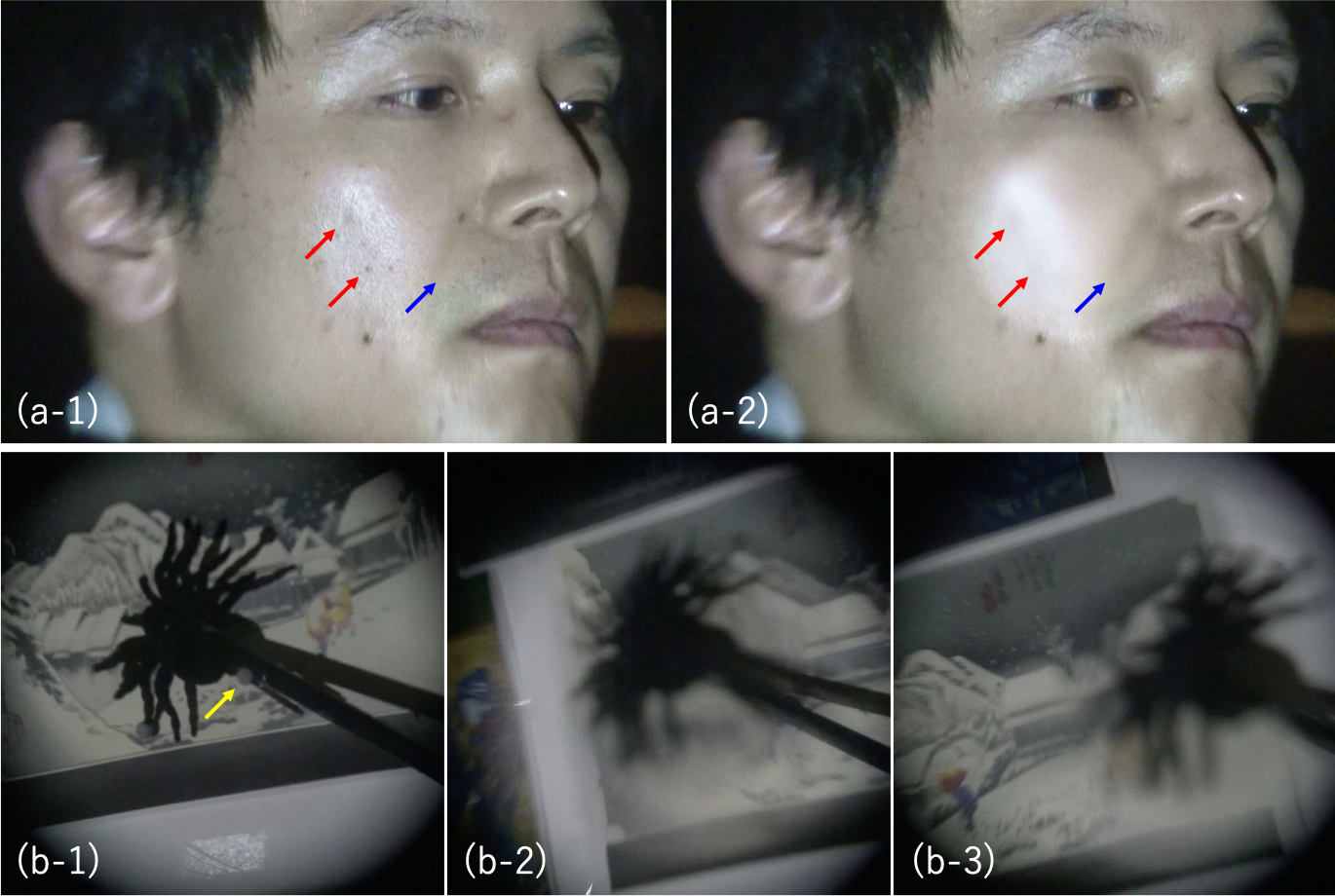}
  \caption{DR applications. (a) Diminishing undesirable signs of facial skin aging by blurring them out. The red and blue arrows indicate blotches and a wrinkle, respectively. \revise{}{(b) Blurring a dynamic object (a toy spider). The yellow arrow indicates a motion capture marker. (b-1) The spider on a black stick is not blurred. (b-2, b-3) The spider appears blurred by the system at different locations.}}
  \label{fig:res_dr}
\end{figure}

As the third application, we implemented a DR system.
This system conceals undesirable signs of facial skin aging (e.g., blotches, pores, and wrinkles) by blurring them out.
Figure \ref{fig:res_dr}(a) and the supplementary video show the results.
\revise{}{We can also apply the proposed system to dynamic objects. Figure \ref{fig:res_dr}(b) shows an example. In this example, the system blurs a moving toy spider. The position of the spider is measured by a motion capture system, and the system interactively moves the blurred area according to the position.}
The same system can be applied to other scenarios.
For example, we can conceal private documents on a shared tabletop in a face-to-face collaboration.
Typical DR methods can replace or fill in an undesired object with its background texture to make it completely invisible.
Our system cannot completely diminish a target; however, it allows a user to see the DR result by their naked eye.
Thus, it can obviously provide the DR result with more reality than typical DR methods that only work on VST displays.

\subsection{DOF enhancement}

% depth enhancement
% 絵画の深度強調
\begin{figure}[t]
  \centering
  \includegraphics[width=0.98\hsize]{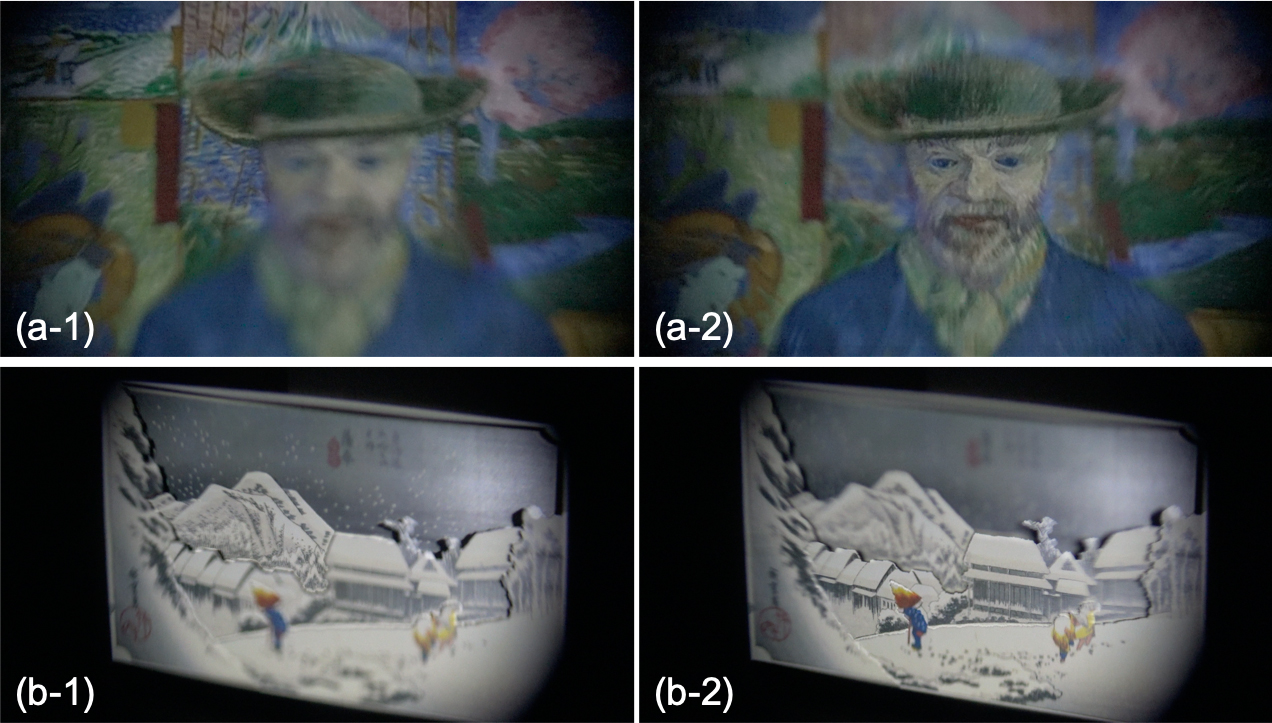}
  \caption{\revise{Emphasizing}{Enhanced} DOF effect for 2D pictures. \revise{}{(a-1, b-1) The foreground object appears blurred. (a-2, b-2) The background object appears blurred.}}
  \label{fig:res_pict}
\end{figure}

\revise{}{For the fourth application, we explored a new vision augmentation scenario.
Because focusing and defocusing affect the perceived 3D structure of a real scene, we believe that our method can affect the depth perception of a real object by blurring it.
We applied our method to \revise{emphasize the DOF}{add an enhanced DOF} effect to 2D pictures.
\revise{}{More specifically, we made either the foreground or background of a picture appear blurred to enhance the perceived depth variation of the picture.}
Figure \ref{fig:res_pict} shows \revise{a result, where it can be perceived that the depth variation of the picture is enhanced.}{a captured result of the application.}
\revise{We asked 10 local participants if they perceive the depth enhancement.}{We showed the portrait picture (Figure \ref{fig:res_pict}(top)) to 10 local participants in two conditions, i.e., with and without the DOF enhancement, and asked in which condition they perceived more depth variation.}
All participants \revise{agreed it}{answered that greater depth variation was perceived with the DOF enhancement}.
Therefore, we confirmed that the proposed method can support this vision augmentation application.}

\section{Discussion}

% a
% 1章後半で述べたように、これまで光学的にはボケは空間コントロールできなかった。それができるようになり、裸眼で見ることのできる様々なアプリケーションが可能になった。
Through the evaluations and application implementations described in Sections \ref{sec:exp} and \ref{sec:app}, we confirmed that depth-independent spatial defocusing is achieved by the proposed IlluminatedFocus technique.
Although previous systems achieved the same technical goal, they applied VST-AR approaches that prevented a user from understanding the context of a real scene and communicating \revise{with others using eye contacts}{with others with good eye contact} because the user's eyes were blocked by a VST display.
The proposed technique solved this problem by allowing the user to see the augmented scene \revise{by the naked eyes}{by nearly naked eyes}.
% シンプルな普通のレンズの公式で、振る舞いが記述できる。
Through the evaluation, we also confirmed that the optical characteristics of the proposed system can be well described by a mathematical model based on the typical thin lens model discussed in Section \ref{sec:method}.
Our design guideline discussed in Section \ref{sec:method:guideline} also worked well in the experiments.
\revise{}{The proposed system is not limited to static objects. It also works for dynamic objects, as shown in Section \ref{sec:app}. In theory, there are no geometric requirements among a viewer, projector, and objects thanks to our design guideline considering the viewer's accommodation, and there is no constraint on the number of objects.}
We believe that an important contribution of this paper is establishing these valid and useful technical foundations, which will allow future researchers and developers to build their own vision augmentation applications on top of the IlluminatedFocus technique.

\revise{}{As discussed in previous computer graphics research \cite{Held:2010:UBA:1731047.1731057}, blur can strongly influence the perceived scale of a rendered scene. A full-size scene would look smaller or a miniature can look bigger. Therefore, inappropriately designed blur would perhaps introduce a false impression. In this paper, we focused on developing a spatial defocusing technology without careful consideration of human perceptual constraints. In future, it would be crucial to investigate this issue and establish a design guideline for vision augmentation applications using the IlluminatedFocus technique for future researchers and developers.}

% limitation
%\begin{figure}[t]
%  \centering
%  \includegraphics[width=0.98\hsize]{pic/res_env.jpg}
%  \caption{The proposed system works under a slightly dim environment light. (left) Experimental setup. (right) Captured appearance where only the middle object was intended to appear blurred.}
%  \label{fig:res_env}
%\end{figure}
\begin{figure}[t]
  \centering
  \includegraphics[width=0.98\hsize]{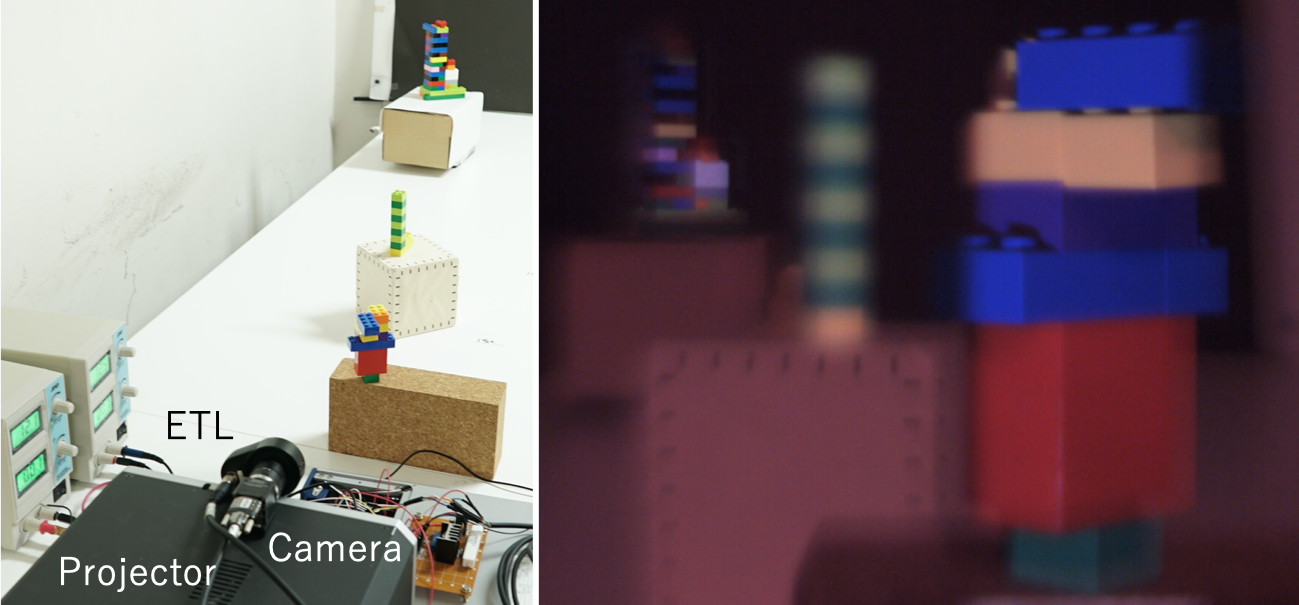}
  \caption{\revise{}{Feasibility test for environment light. Left: experimental setup. Right:} captured appearance\revise{ in the proposed system}{} under a slightly dim environment light (40 lux). Only the middle object was intended to appear blurred.}
  \label{fig:res_env}
\end{figure}

A current limitation of the proposed system is its small lens aperture.
The aperture of the current system is 16 mm, which limits the FOV of a user.
However, the ETL industry is an emerging field and ETL's characteristics \revise{is}{are} being improved rapidly.
For example, ETLs have been getting thinner and lighter.
Recently, it was announced that a new ETL provides an aperture that is almost two times larger (30 mm) than the current one \cite{Padmanabaneaav6187}.
Therefore, we believe that the form factor of our system can be similar to typical eyeglasses in near future, which solves the FOV problem.
Another limitation is that we assume to use our technique in a dark environment where only the projector illuminates the scene.
We conducted a simple experiment to check how an environment light affected the spatial defocusing result in our system.
Figure \ref{fig:res_env} shows the result under slightly dim lighting (40 lux). From this result, we confirmed that the system appeared to work under environmental lighting (see Figure \ref{fig:teaser}(b) for the experimental setup).
However, in theory, a user could see both the focused and defocused appearances of the scene.
Investigating the visual perception of the user in this condition is interesting future work.
Another more technical future direction is to build an environment where multiple projectors cooperatively illuminate the real scene such as in a previous study \cite{Fender:2017:MRO:3132272.3134117}, resulting in an environment is not dark.

\begin{figure}[t]
  \centering
  \includegraphics[width=0.98\hsize]{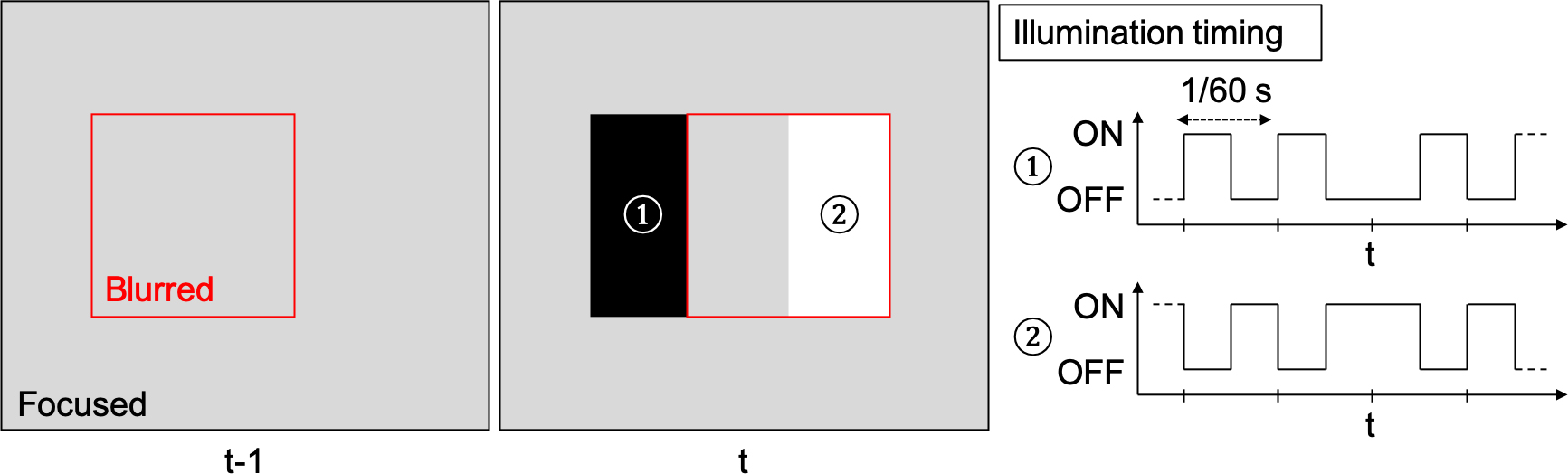}
  \caption{\revise{}{Edges of a blurred area become visible when it moves fast.}}
  \label{fig:fast_mov}
\end{figure}

\revise{}{The current system does not assume fast movement of a blurred area. Fast movement makes the edge of the blurred area visible. Assume a simple case where a squared blurred area moves to the right (Figure \ref{fig:fast_mov}). In addition, assume we illuminate the blurred area in the first half of 1/60 s and the focused area in the second half. In this situation, the right edge is illuminated for 1/60 s when it moves and the left edge is not illuminated for 1/60 s. Consequently, the edges are visible to a user. We apply blending to the edges to alleviate the visible seam as described in Section \ref{sec:exp:seam}, which suppresses this effect when the movement is sufficiently slow \cite{4538845}. However, if the movement is too fast, the edge becomes conspicuous to the viewer. This problem occurs in the DR application for human skin, as shown in the supplementary video. In future, we intend to modify our design guideline to solve this problem.}

% 明るさとfocus accuracyのトレードオフ
% 狙いのジオプトリーのみ提示したい => 井澤scientific reports
The size of the blur circle in the proposed system is slightly larger than that in a normal lens system, as shown in Section \ref{sec:exp:size}.
This is due to the integration of blur circles of different sizes within the illumination period (i.e., 1 ms in our experiment).
To generate the optical power more accurately, we need to illuminate a scene for a shorter period.
DLP projectors are capable of illuminating a real scene at a much higher frame rate ($>$ 40k fps).
However, the higher frame rate results in an appearance that is too dark for a user to perceive.
Therefore, there is a tradeoff between the accurate optical power generation and the brightness of the scene.
We can address the tradeoff by controlling the ETL more flexibly.
Currently, we apply a simple sinusoidal wave as the input signal to the ETL; consequently, its optical power stays at the target power for a very short period.
We can increase the period by applying a staircase modulation \cite{izawa}.
This allows a projector to illuminate the real scene for a longer period during which the size of the blur circle does not change.

\section{Conclusion}

% Aiming at realizing novel vision augmentation experiences,
This paper presented the IlluminatedFocus technique by which spatial defocusing of real-world appearances is achieved regardless of distances from users' eyes to observed real objects.
The core of the technique is the combination of focal sweep eyeglasses using ETLs and a synchronized high-speed projector as illumination for a real scene.
We described the technical details involved in achieving the spatial defocusing including the mathematical model of the blur range, the design guideline of the illumination timing and the focal sweep range, and the technique to alleviate a visible seam between focused and blurred regions.
Through the experiments, we confirmed the feasibility of our proposal for vision augmentation applications.
As future works, we will investigate more robust and flexible techniques to use the method under more relaxed conditions, such as under environment lighting.

\acknowledgments{
This work was supported by JSPS KAKENHI grant number JP17H04691 and JST, PRESTO Grant Number JPMJPR19J2, Japan.}

\bibliographystyle{abbrv-doi}

\bibliography{template}
\end{document}